\documentclass[fleqn,usenatbib]{mnras}

\usepackage{mathptmx}
\usepackage{times}

\usepackage[T1]{fontenc}

\DeclareRobustCommand{\VAN}[3]{#2}
\let\VANthebibliography\thebibliography
\def\thebibliography{\DeclareRobustCommand{\VAN}[3]{##3}\VANthebibliography}

\usepackage{graphicx}	
\usepackage{amsmath}	
\usepackage{amssymb}
\usepackage{wasysym}

\newcommand{\Mpc}{\rm\; Mpc}
\newcommand{\kpc}{\rm\; kpc}

\newcommand{\km}{\rm\; km}

\newcommand{\cm}{\rm\; cm}
\newcommand{\mum}{\hbox{$\rm\; \mu m\,$}}
\newcommand{\pix}{\rm\; pixel}

\newcommand{\ppix}{\hbox{$\pix^{-1}\,$}}

\newcommand{\yr}{\rm\; yr}
\newcommand{\Gyr}{\rm\; Gyr}
\newcommand{\Myr}{\rm\; Myr}
\newcommand{\s}{\rm\; s}

\newcommand{\ks}{\rm\; ks}
\newcommand{\mus}{\hbox{$\rm\; \mu s\,$}}

\newcommand{\GHz}{\rm\; GHz}

\newcommand{\Hz}{\rm\; Hz}

\newcommand{\Msun}{\hbox{$\rm\thinspace M_{\odot}$}}

\newcommand{\Msunpyr}{\hbox{$\Msun\yr^{-1}\,$}}

\newcommand{\keV}{\rm\; keV}
\newcommand{\eV}{\rm\; eV}
\newcommand{\erg}{\rm\; erg}

\newcommand{\W}{\rm\; W}

\newcommand{\ergpcmcu}{\hbox{$\erg\cm^{-3}\,$}}
\newcommand{\ergpcmsqps}{\hbox{$\erg\cm^{-2}\s^{-1}\,$}}

\newcommand{\ergps}{\hbox{$\erg\s^{-1}\,$}}

\newcommand{\keVcmsq}{\hbox{$\keV\cm^{2}\,$}}
\newcommand{\WpHzpsr}{\hbox{$\W\pHz\psr\,$}}

\newcommand{\ctrate}{\hbox{$\rm\thinspace cts\ps$}}
\newcommand{\surbri}{\hbox{$\rm\thinspace cts\pcmsq\ps\pasecsq$}}
\newcommand{\expmapcorr}{\hbox{$\rm\thinspace cts\pcmsq\ps\ppix$}}
\newcommand{\kmps}{\hbox{$\km\s^{-1}\,$}}

\newcommand{\kmpspMpc}{\hbox{$\kmps\Mpc^{-1}\,$}}

\newcommand{\Lsun}{\hbox{$\rm\thinspace L_{\odot}$}}

\newcommand{\Zsun}{\hbox{$\thinspace \mathrm{Z}_{\odot}$}}

\newcommand{\ionpar}{\hbox{$\erg\cm\ps\,$}}

\newcommand{\amin}{\rm\; arcmin}

\newcommand{\asec}{\rm\; arcsec}

\newcommand{\sr}{\rm\; sr}

\newcommand{\sqremm}{\hbox{$\cm^{-5/2}\,$}}

\newcommand{\pseudoP}{\hbox{$\keV\sqremm\asec^{-2}\,$}}

\newcommand{\pcmsq}{\hbox{$\cm^{-2}\,$}}
\newcommand{\pcmcu}{\hbox{$\cm^{-3}\,$}}

\newcommand{\ps}{\hbox{$\s^{-1}\,$}}
\newcommand{\pHz}{\hbox{$\Hz^{-1}\,$}}

\newcommand{\psr}{\hbox{$\sr^{-1}\,$}}
\newcommand{\pasecsq}{\hbox{$\asec^{-2}\,$}}

\begin{document}

\title[Cooling flow around H1821+643]{A cooling flow around the low-redshift quasar H1821+643}
\author[H.R. Russell et al.]  
       {\parbox[]{7.in}{H.~R. Russell$^{1}$\thanks{E-mail: helen.russell@nottingham.ac.uk},
           P.~E.~J. Nulsen$^{2,3}$,
           A.~C. Fabian$^4$,
           T.~E. Braben$^{1}$,
           W.~N. Brandt$^{5,6,7}$,
           L. Clews$^{1,8}$,
           M. McDonald$^{9,10}$,
           C.~S. Reynolds$^{11,12}$,
           J.~S. Sanders$^{13}$
           and S. Veilleux$^{11,12}$
           \\   
     \footnotesize
     $^1$ School of Physics \& Astronomy, University of Nottingham, University Park, Nottingham NG7 2RD, UK\\
     $^2$ Center for Astrophysics | Harvard \& Smithsonian, 60 Garden Street, Cambridge, MA 02138, USA\\
     $^3$ ICRAR, University of Western Australia, 35 Stirling Hwy, Crawley, WA 6009, Australia\\
     $^4$ Institute of Astronomy, Madingley Road, Cambridge CB3 0HA, UK\\
     $^5$ Department of Astronomy \& Astrophysics, 525 Davey Lab, The Pennsylvania State University, University Park, PA 16802, USA\\
     $^6$ Institute for Gravitation and the Cosmos, The Pennsylvania State University, University Park, PA 16802, USA\\
     $^7$ Department of Physics, 104 Davey Lab, The Pennsylvania State University, University Park, PA 16802, USA\\
     $^8$ School of Physical Sciences, The Open University, Walton Hall, Milton Keynes, MK7 6AA, UK\\
     $^9$ Department of Physics, Massachusetts Institute of Technology, Cambridge, MA 02139, USA\\
     $^{10}$ Kavli Institute for Astrophysics and Space Research, Massachusetts Institute of Technology, 77 Massachusetts Avenue, Cambridge, MA 02139, USA\\
     $^{11}$ Department of Astronomy, University of Maryland, College Park, MD 20742-2421, USA\\
     $^{12}$ Joint Space-Science Institute (JSI), College Park, MD 20742-2421, USA\\
     $^{13}$ Max-Planck-Institut f{\"u}r extraterrestrische Physik, Gie{\ss}enbachstra{\ss}e 1, D-85748, Garching, Germany\\
  }
}
    
\maketitle

\begin{abstract}
H1821+643 is the nearest quasar hosted by a galaxy cluster.  The energy output by the quasar, in the form of intense radiation and radio jets, is captured by the surrounding hot atmosphere.  Here we present a new deep \textit{Chandra} observation of H1821+643 and extract the hot gas properties into the region where Compton cooling by the quasar radiation is expected to dominate.  Using detailed simulations to subtract the quasar light, we show that the soft-band surface brightness of the hot atmosphere increases rapidly by a factor of $\sim30$ within the central $\sim10\kpc$.  The gas temperature drops precipitously to $<0.4\keV$ and the density increases by over an order of magnitude.  The remarkably low metallicity here is likely due to photo-ionization by the quasar emission.  The variations in temperature and density are consistent with hydrostatic compression of the hot atmosphere.  The extended soft-band peak cannot be explained by an undersubtraction of the quasar or scattered quasar light and is instead due to thermal ISM.  The radiative cooling time of the gas falls to only $12\pm1\Myr$, below the free fall time, and we resolve the sonic radius.  H1821+643 is therefore embedded in a cooling flow with a mass deposition rate of up to $3000\Msunpyr$.  Multi-wavelength observations probing the star formation rate and cold gas mass are consistent with a cooling flow.  We show that the cooling flow extends to much larger radii than can be explained by Compton cooling. Instead, the AGN appears to be underheating the core of this cluster.
\end{abstract}


\section{Introduction}
\label{sec:intro}

\begin{table*}
\begin{minipage}{\textwidth}
\caption{}
\begin{center}
\begin{tabular}{l c c c l c c c}
\hline
Date & Obs. ID & Aim point & Exposure & Date & Obs. ID & Aim point & Exposure\\
 & & & (ks) &  & & & (ks) \\
\hline
2019 October 7 & 22105 & S3 & 36.3 & 2020 July 18 & 23319 & S3 & 20.4   \\ 
2019 October 23 & 22106 & S3 & 45.4 & 2020 July 19 & 23239 & S3 & 39.1   \\ 
2019 December 28 & 23054 & S3 & 41.8 &     2020 August 7 & 22104 & S3 & 20.9   \\ 
2020 April 10 & 21559 & S3 & 25.5 &        2020 August 9 & 23339 & S3 & 14.6   \\ 
2020 April 10 & 23211 & S3 & 18.2 &        2020 August 22 & 23053 & S3 & 22.7   \\ 
2020 April 20 & 21558 & S3 & 42.7 &        2020 September 16 & 22103 & S3 & 13.7   \\ 
2020 May 8 & 21561 & S3 & 24.1 &           2020 September 18 & 22109 & S3 & 22.7   \\ 
2020 May 10 & 23240 & S3 & 22.7 &          2020 September 19 & 24612 & S3 & 27.6   \\ 
2020 May 28 & 22108 & S3 & 13.7 &          2020 September 24 & 24639 & S3 & 30.9   \\ 
2020 June 20 & 22107 & S3 & 34.4 &         2020 September 26 & 24641 & S3 & 22.7   \\ 
2020 July 15 & 21560 & S3 & 32.7 &         2020 September 27 & 24661 & S3 & 9.1   \\ 
\hline
\end{tabular}
\end{center}
\label{tab:obs}
\end{minipage}
\end{table*}

Accretion onto supermassive black holes (SMBHs) powers the intense
radiation and winds observed in distant quasars and spectacular
relativistic jets in radio galaxies that can reach far beyond the host
galaxy.  This energy input is now understood to be the key mechanism
in structure formation that truncates the growth of massive galaxies and
suppresses cooling flows at the centres of rich galaxy clusters
(e.g. \citealt{Croton06,Bower06,Hopkins06} and for a review see
\citealt{Fabian12}). The clearest observational evidence for this
mechanism is found in nearby galaxy clusters, where the energy output
by the central AGN is imprinted on the hot atmosphere. In these
systems, radio jets launched by the SMBH carve out huge cavities and
drive a series of weak shocks and sound waves into the hot gas
(e.g. \citealt{McNamara00,FabianPer00,FabianPer06} and for a review see
\citealt{McNamaraNulsen07}). The radio jets and lobes expand against
the pressure of the hot atmosphere; therefore the minimum energy input
is given by $4pV$, where $p$ is the gas pressure and the cavity
volume $V$ is based on the extent on the sky
(e.g. \citealt{Churazov02}). The captured energy replaces the
substantial radiative losses from the cluster’s atmosphere, which
would otherwise cool rapidly and flood the central galaxy with cold
gas (a cooling flow, \citealt{Fabian94}). This radio or maintenance-mode of feedback
is thought to suppress gas cooling and star formation in massive
galaxies and clusters at late times in the universe.

At high redshift, higher accretion rates onto these SMBHs fuelled a
radiatively efficient quasar-mode of feedback. In this mode, intense
radiation and high velocity winds generated close to the quasar
shock-heat and expel the surrounding gas and prevent further gas inflow 
(e.g. \citealt{DiMatteo05,King15}). Quasar-mode feedback was probably
most effective and prevalent at the height of quasar activity at
$z\sim2-3$, when galaxies were most gas rich and intensively forming
stars. By driving outflows and depleting the fuel supply, quasar-mode
feedback would produce the observed decline in activity from both
black holes and star formation and push galaxies onto the
$M_{\mathrm{BH}}-\sigma$ relation.  However, detailed studies of the
impact of these powerful quasars on hot atmospheres have so far
been limited by the lack of suitable low-redshift analogs. Only two
are known in the local universe ($z<0.5$): H1821+643 ($z=0.299$;
\citealt{Russell10}) and CL09104+4109, which is highly obscured
($z=0.44$; \citealt{OSullivan12}).

The massive galaxy cluster surrounding the quasar H1821+643 is
optically rich and has an X-ray bright, strong cool core. The quasar
has a bolometric luminosity of $\sim10^{47}\ergps$ and this nuclear
emission dominates over that from the hot gas within the central few
arcsec (\citealt{Fang02}). A deep VLA observation revealed that,
although classified as a radio-quiet quasar (based on the $5\GHz$
luminosity of $10^{23.9}\WpHzpsr$), H1821+643 hosts a giant $300\kpc$
FR I radio source (\citealt{Blundell01}). Previous Chandra imaging
observations suggest complex interplay between the AGN activity and
the surrounding cluster gas (\citealt{Russell10}). Bright arms of cool
gas extend along the radio lobes and around the rim of an X-ray
cavity. Using detailed PSF simulations to subtract the quasar
emission, \citet{Russell10} extracted the underlying temperature and
density profiles of the hot atmosphere. The gas temperature drops
steadily from $10\keV$ at $\sim80\kpc$ to $2\keV$ at
$20\kpc$. When compared to other strong cool core clusters, the
resulting entropy profile for H1821+643 has a significantly steeper
decline within the central $\sim80\kpc$ (\citealt{Walker14}). The
quasar may have a marked cooling effect on the surrounding hot gas,
likely through Compton scattering, which would also boost the
accretion rate onto the SMBH and extend this burst of quasar activity
(\citealt{Fabian90}). However, heavy pileup of the quasar emission in
the previous Chandra observations ($>80\%$ of photons affected) ultimately prevented us
from extracting the underlying hot gas properties into the region
where Compton cooling is expected to dominate.

\begin{figure*}
  \begin{minipage}{\textwidth}
    \centering
    \includegraphics[width=0.45\columnwidth]{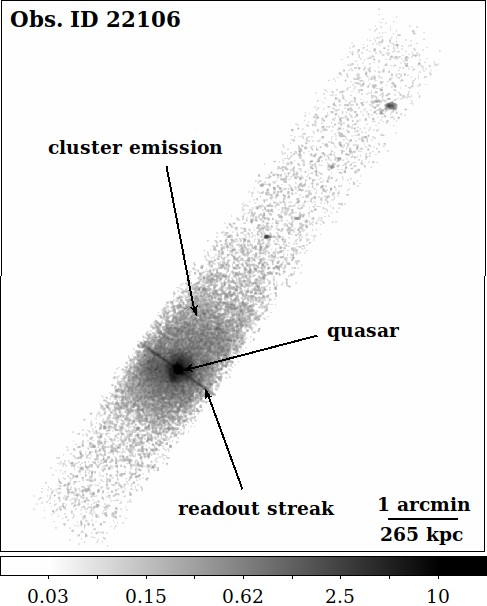}
    \includegraphics[width=0.45\columnwidth]{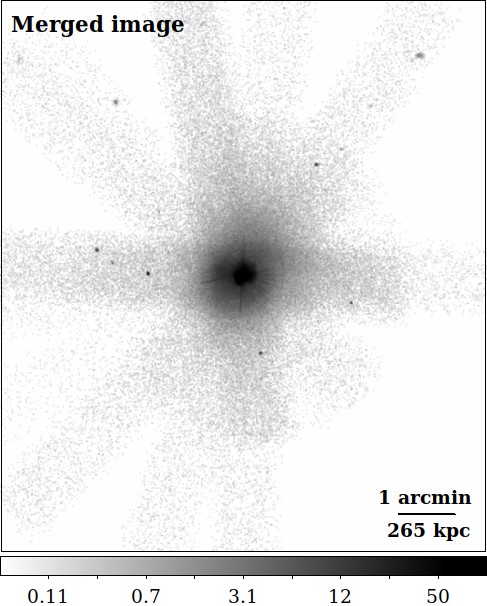}
    \caption{Left: Image for a single \textit{Chandra} observation of H1821+643 showing the 1/8 sub-array and readout streak from the bright quasar (obs. ID 22106).  Right: Image showing all datasets merged.  The range of roll angles for the separate observations changes the orientation of the CCD and readout streak.  Images have units of counts, cover the energy range $0.5-7\keV$ and were smoothed with a 2D Gaussian with $\sigma=1.5\asec$.}
    \label{fig:rawimage}
  \end{minipage}
\end{figure*}

Here we present a new deep \textit{Chandra} observation of H1821+643
taken with a short frame-time of $0.4\s$ to minimise pileup and allow
us to measure the temperature and density of the hot gas within the
central $20\kpc$.  In sections \ref{sec:data} and \ref{sec:PSF}, we discuss the data reduction, pileup and the detailed simulation and subtraction of the quasar PSF.  Section \ref{sec:image} examines the morphology of the underlying hot atmosphere and section \ref{sec:spec} maps the gas properties using spatially-resolved spectroscopy.  Finally, in section \ref{sec:disc}, we discuss the evidence for and origin of the cooling flow at the centre of H1821+643.  We assume $H_0=70\kmpspMpc$, $\Omega_{\mathrm{m}}=0.3$ and
$\Omega_\Lambda=0.7$, translating to a scale of $4.4\kpc$ per arcsec
at the redshift $z=0.296$ of H1821+643 (\citealt{Monroe16}).  All errors are $1\sigma$ unless otherwise noted.


\section{Data reduction}
\label{sec:data}

H1821+643 was observed by \textit{Chandra} with the ACIS-S detector
for a total of $582\ks$ split over 22 separate observations between
October 2019 and September 2020 (Fig. \ref{fig:rawimage}).  The bright
quasar was positioned $1\amin$ off-axis to ensure the field of view
covered the galaxy cluster core.  \textit{Chandra} does not have a shutter so
quasar photons arriving during CCD readout produce a clear
readout streak across the detector.  All observations were taken
with a 1/8 sub-array and reduced frame time of $0.4\s$ to minimise
pileup.  Pileup occurs when multiple photons arrive in the same pixel
within a single ACIS frame integration time (typically $3.1\s$).
These photons are detected as a single event of higher energy and with
an extended charge cloud distribution.  An event with an extended
charge cloud distribution is more likely to be flagged as a background
event and excluded from analysis.  Pileup therefore hardens the source
spectrum and reduces the detected flux.  Pileup can be reduced by
reducing the frame time, however, this requires faster readout so the
CCD area must be decreased by using a sub-array.

Using PIMMS (\citealt{Mukai93}), we estimated that $\sim25\%$ of quasar
photons are piled up in the new observations.  Whilst far below the
$>80\%$ pile up predicted for a frame time of $3.1\s$, this is still
sufficient to reduce the measured count rate from $1.8\ctrate$ to
$1.0\ctrate$ and produce a coincident peak in bad event grades.  Based
on the ratio of bad to good grades, pileup predominantly affects the
very centre of the quasar's PSF within $1\asec$.  We note that mild
pileup ($<5\%$) is also found from $1-2\asec$, which is considered when
analysing the radial profiles for the extended emission.

The datasets were reduced and analysed with \textsc{ciao} version 4.14
and \textsc{caldb} version 4.9.8 provided by the \textit{Chandra}
X-ray Center (\citealt{Fruscione06}).  The latest gain and charge
transfer inefficiency correction were applied and the event files were
filtered for bad grades.  For this analysis, the improved background
screening available from \textsc{VFAINT} mode was not applied to
minimise loss of piled up photons.  Background light curves were
extracted from the outskirts of the CCD and filtered with the
\textsc{lc\_clean} script to remove periods impacted by flares.  No
major flares were found in the individual background light curves.  The
light curves were also compared to ensure no flares were missed in the
shorter observations.  No flare periods were identified.  The
observations used and their corresponding exposure times are listed in
Table \ref{tab:obs}.

The earlier $90\ks$ observation of H1821+643 taken with ACIS-S in 2008
was severely piled up ($>80\%$) within the region of interest here.
Piled up events exceeded the upper energy cutoff and were not
telemetered to the ground, which resulted in an apparent hole at the
centre of the quasar PSF.  This observation was focused on the
surrounding galaxy cluster rather than the immediate environment of
the quasar and therefore did not use a sub-array
(\citealt{Russell10}).  We did not utilize this additional dataset in
this study except to evaluate systematic uncertainty due to the
contaminant correction.  \textit{Chandra's} ACIS instrument has
decreasing soft-band effective area over time due to the deposition of
a contaminant on the optical blocking filters.  The rate of
accumulation is varying over time and the calibration must be
regularly updated (\citealt{Plucinsky18}).  We therefore compared the
best-fit parameters for the new and 2008 observations for identical
regions at large radius to check the accuracy of the contaminant
correction.  We found no systematic differences.

The cluster emission extended over the full sub-array so blank sky
backgrounds were generated for each observation.  Each background data
set was reprocessed as above, reprojected to the appropriate sky
position and normalized to match the observed count rate in the
$9.5-12\keV$ energy band.  We note that this study focused on the
luminous extended emission immediately around the quasar therefore the
background subtraction was not particularly significant and the blank
sky backgrounds were sufficiently well-matched to the observation.

Sub-pixel event repositioning was used to determine the quasar
centroid for image and profile alignment across the separate
observations (\citealt{Li04}).  For each observation, an image of the
quasar was generated with a spatial resolution 10 times finer than the
native resolution of $0.495\asec$.  The quasar centre was then
determined with 2D image fitting in \textsc{sherpa}
(\citealt{Freeman01}).  The best-fit centroid for each dataset was
used throughout to align images and position extraction regions for radial profiles and
spectra.  Pileup and, to a lesser extent, the presence of extended
emission will distort the PSF shape.  However, this effect will be
comparable for all datasets and was found to be negligible in a
similar analysis of M87, which has more complex circumnuclear emission
(\citealt{Russell18}).  Due to the sub-array's limited field of view,
additional point sources were only detected in a few observations
taken at favourable roll angles.  This limitation on the alignment, in
combination with PSF asymmetries at $1\amin$ off-axis, prevented us
from analysing sectors on small scales around the quasar.

\section{Quasar PSF subtraction}
\label{sec:PSF}

The quasar PSF emission dominates over the extended emission from the
hot atmosphere to a radius of $4\asec$ ($18\kpc$) for the energy band
$0.5-7\keV$.  The quasar is so bright that the PSF emission only drops
below $10\%$ of the total $0.5-7\keV$ flux beyond a radius of
$20\asec$ ($90\kpc$).  The \textit{Chandra} PSF is broader at high
energies compared to low energies so the impact can be even greater in
a hard energy band.  The quasar PSF has a significantly harder
spectrum than the thermal emission so this contribution must be
carefully modelled to large radii to remove systematic increases in
both normalization and measured gas temperature.  It is not possible
to cleanly spectrally separate the spatially coincident quasar and
extended emission (\citealt{Russell10}).  Instead, we must determine
the PSF contribution at each radius, fix the parameters of an
appropriate spectral model that accounts for this contribution and
then fit a thermal model to the residual emission from the hot
atmosphere.

A detailed model of the HRMA PSF was produced before launch and
calibrated with in-flight data (\citealt{Jerius00}).  This model is
continually reviewed and updated by the \textit{Chandra} X-ray
Center\footnote{See
https://cxc.harvard.edu/ciao/PSFs/psf\_central.html}.  Given the
energy dependence of the PSF, the principle input to this model is the
incident quasar spectrum.  


\subsection{Quasar spectrum}
\label{sec:quasarspec}

\begin{figure}
  \centering
  \includegraphics[width=0.98\columnwidth]{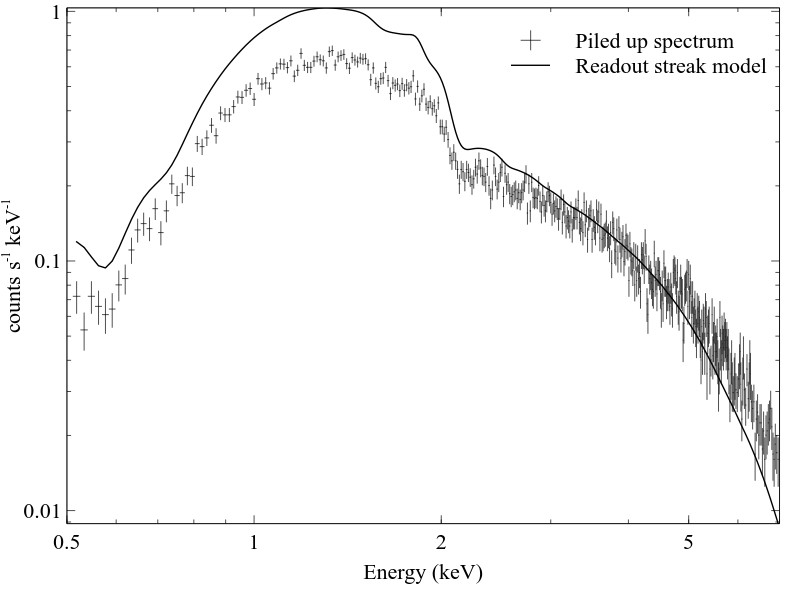}
  \caption{Quasar spectrum extracted from obs. ID 21558 using a circular region with radius $1.5\asec$.  Approximately $25\%$ of the incident photons were piled up in the spectrum.  The spectral model determined from the readout streak spectrum is shown by the solid line.  This model is unaffected by pileup.  The difference between the model and the piled up spectrum shows how pileup hardens the observed spectrum and reduces the flux.}
  \label{fig:pileupspec}
\end{figure}

Pileup significantly distorts the observed spectrum of the quasar
H1821+643.  Fig. \ref{fig:pileupspec} shows the modestly piled up
spectrum extracted from a circular region centred on the quasar with a
radius of $1.5\asec$.  The model illustrates the incident spectrum and
demonstrates that when multiple soft photons arrived in a single
frame-time these were recorded as a single hard photon.  The piled up
spectrum has a significant deficit of soft photons and an excess of
hard photons.  The piled up spectrum can be fitted with a statistical
model of pileup (\citealt{Davis01}) to recover the incident spectrum.
However, the solutions were found to be particularly degenerate for
this dataset and the best-fit result predicted a far higher level of
pileup than was consistent with the fraction of bad grades or readout
streak brightness.

\begin{figure}
  \centering
  \includegraphics[width=0.98\columnwidth]{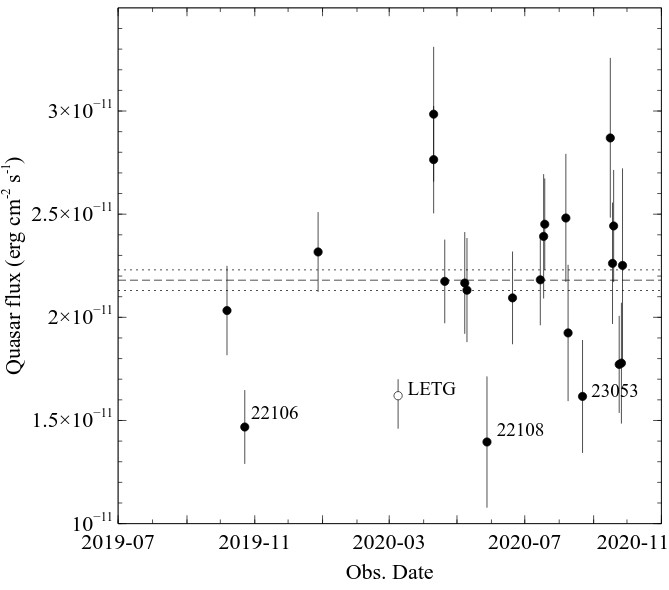}
  \caption{Unabsorbed quasar flux from the power-law component determined for each readout streak spectrum and plotted as a function of observation date.  Observations with fluxes significantly different from the average are labelled.  The quasar flux when fitting all observations together is shown by the dashed line with uncertainties shown by the dotted lines.  The LETG data point is also shown for comparison.}
  \label{fig:variability}
\end{figure}

Other techniques to extract the incident spectrum include analysing
the PSF wings (e.g. \citealt{Evans04}, \citealt{Hardcastle06},
\citealt{Mingo11}) and using only the best event grade to select single photon arrivals (grade 0, e.g. \citealt{Miller17}).  However,
the former suffered from degeneracies between the spectrum of the PSF
wings and the extended hot atmosphere emission.  The latter produces a
spectrum free of pile up distortions but cannot recover the flux of
the incident spectrum because the ratio of grade 0 events to total
incident photons remains unknown.

We therefore extracted the quasar's incident spectrum from the readout
streak (see e.g. \citealt{Russell10}).  For each observation, ACIS
accumulates photons over the frame-time of $0.4\s$ and then reads out
the frame at a parallel transfer rate of $40\mus$.  For the 1/8
sub-array, only 128 rows are read, which takes a total of $0.00512\s$.
Unlike an optical telescope, the typically very low count rate of
X-ray sources ensures that a negligible number of photons arrive
during readout and a shutter is not required.  In the case of very
bright point sources, such as H1821+643, photons do arrive during the
CCD readout and appear distributed along the entire row, which
creates a continuous streak.  As the effective frame-time is then much
shorter ($40\mus$), these photons are not piled up.

We extracted a spectrum in the energy range $0.5-7\keV$ from two
narrow regions positioned over the readout streak on either side of
the quasar.  Each region was $20-30$ pixels long and $6$ pixels wide.
Background cluster emission was subtracted by extracting spectra from
closely neighbouring background regions either side of the readout
streak.  Appropriate responses were also generated.  The exposure time
of each readout streak spectrum was calculated by multiplying together
the number of rows, the number of frames and the transfer rate of
$40\mus$.  For the observations of H1821+643, the readout streak
exposure times ranged from $45.7$ to $227.1\s$ with a total of $2.9\ks$.

The readout streak spectra were fit together and separately in the
X-ray spectral fitting package \textsc{xspec} (version 12.12.1,
\citealt{Arnaud96}) using an absorbed power-law model.  Two absorption
components were used: the Galactic absorption component was fixed to
the measured value from the LAB H\textsc{i} survey
($N_{\mathrm{H}}=3.5\times10^{20}\pcmsq$, \citealt{Kalberla05}) and
the intrinsic absorption component was left free.  When fit together,
the best-fit photon index $\Gamma=2.13\pm0.06$, the intrinsic
absorption $N_{\mathrm{H,z}}<0.025\times10^{22}\pcmsq$ and the
unabsorbed flux $F=2.18\pm0.05\times10^{-11}\ergpcmsqps$ ($0.5-7\keV$,
and $F=1.22\pm0.07\times10^{-11}\ergpcmsqps$ for $2-10\keV$).  This best-fit model is shown in Fig. \ref{fig:pileupspec}.  When
fit separately, the best-fit photon index and intrinsic absorption for
each observation are consistent within their respective uncertainties.
The quasar flux shows evidence for modest variability by up to a
factor of 2 (Fig. \ref{fig:variability}).

\begin{figure}
\centering
\includegraphics[width=0.98\columnwidth]{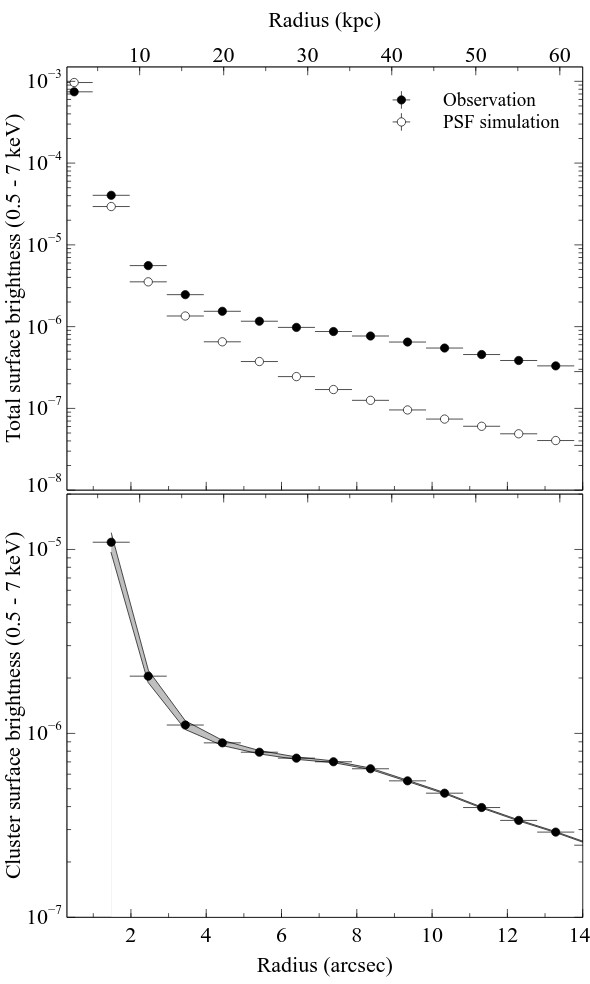}
\caption{Top: Surface brightness profiles ($\surbri$) for the
  observation and the PSF simulation in the energy range $0.5-7\keV$.
  Bottom: Residual surface brightness profile tracing the extended
  emission of the hot atmosphere (PSF simulation subtracted).
  Uncertainties from the best-fit quasar spectral model are shown by
  the shaded band.  The PSF simulation does not include pileup, therefore the predicted PSF flux exceeds the observed flux for the $1-2\asec$ datapoint.}
\label{fig:sbprofilefull}
\end{figure}

Given the apparent lack of variability in the photon index and
intrinsic absorption, we used an archival HRC-LETG observation of
H1821+643 (obs. ID 22958, PI Kraft) that was taken in March 2020,
roughly in the middle of our ACIS program, to verify the best-fit
results.  The LETG spectra were taken from the \textit{Chandra}
grating data archive and catalog (TGCat, \citealt{Huenemoerder11}) and
fit with the same absorbed powerlaw model.  The best-fit photon index
$\Gamma=2.11^{+0.07}_{-0.06}$ is consistent with the readout streak
spectrum result within the uncertainties and the intrinsic absorption
is a comparable but more tightly constrained upper limit
($N_{\mathrm{H,z}}<0.009\times10^{22}\pcmsq$).  The LETG unabsorbed
flux $F=1.62^{+0.08}_{-0.16}\times10^{-11}\ergpcmsqps$ for $0.5-7\keV$
is also consistent within the range of variability spanned by the ACIS
observations.  We therefore proceed with the quasar model spectrum
given by the best-fit parameters from the readout streak spectra with
a jointly fit $\Gamma=2.13\pm0.06$ and zero intrinsic absorption and
power-law normalization determined separately for each observation.


The best-fit photon index is significantly steeper than
$\Gamma=1.761^{+0.047}_{-0.052}$ measured with \textit{Chandra} HEG
and MEG spectra in 2002 (\citealt{Fang02}).  Using a readout streak
spectrum, \citet{Russell10} found a consistent $\Gamma$ and measured an
unabsorbed flux of $1.45^{+0.04}_{-0.03}\times10^{-11}\ergpcmsqps$ for
an energy range of $2-10\keV$.  The flux is therefore consistent with
the current level of variability spanned by the new ACIS observations.

\subsection{PSF simulations}
\label{sec:PSFsim}

\begin{figure*}
  \begin{minipage}{\textwidth}
    \centering
    \includegraphics[width=0.45\columnwidth]{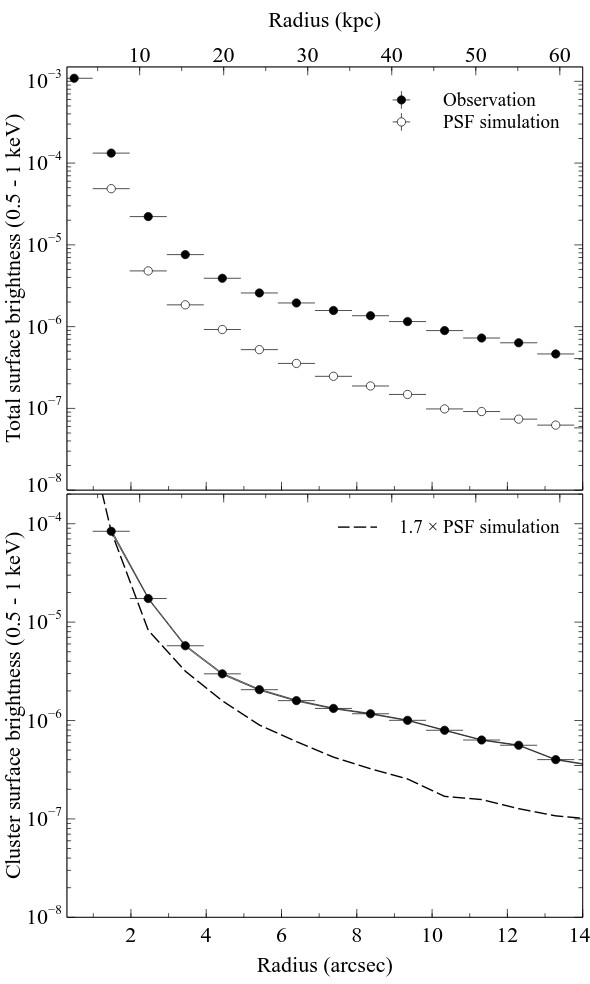}
    \includegraphics[width=0.45\columnwidth]{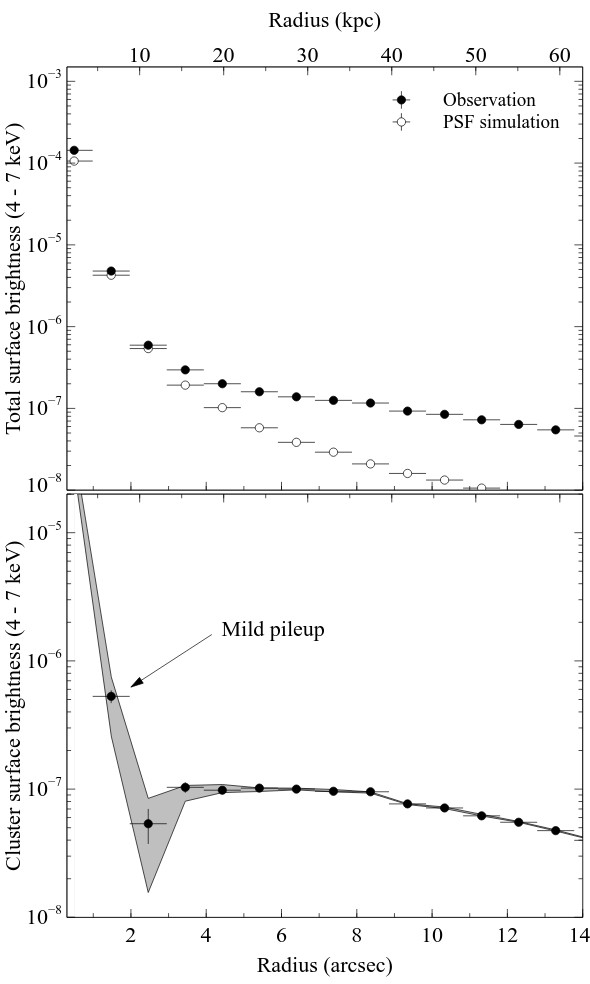}
    \caption{The soft-band surface brightness profile ($0.5-1\keV$) is dominated by emission from the centre of the hot atmosphere (left).  The hard-band surface brightness profile ($4-7\keV$) is dominated by quasar PSF emission and the projected outer, hotter layers of the cluster atmosphere.  Top panels show the observed profiles and the quasar PSF simulation.  The bottom panels show the residual profiles where the PSF simulation has been subtracted.  Units are $\surbri$.  The hard-band profile is free of pileup beyond $2\asec$ radius and shows no strong residuals on arcsec scales from the quasar PSF.  In contrast, the soft-band profile shows a clear central spike that is broader than the PSF.}
    \label{fig:sbprofiles}
  \end{minipage}
\end{figure*}

The incident quasar spectral model was input to the \textit{Chandra}
Ray Tracer (ChaRT, \citealt{Carter03}) to generate simulations of the
PSF at the $1\amin$ off-axis angle.  Multiple realizations were
combined to ensure a full sampling of the range of optical paths in
the HRMA and reduce statistical errors.  \textsc{marx} version 5.5.1
(\citealt{Davis12}) was then used to project the simulated rays onto
the ACIS-S detector and produce a simulated events file for an
observation of the quasar.  Pileup effects were not included.  Whilst
\textsc{marx} can simulate the effects of pileup, the extended
emission increases the level of pileup in the observation and this
cannot be captured by a simulation of the quasar alone.
Instead, we consider the impact of mild pileup ($<5$\%) in the $1-2\asec$ annulus on our analysis.  Simulated
quasar surface brightness profiles, spectra and responses were then
extracted for the same spatial regions analysed in the observation.
The flux was scaled by the relative flux of each readout streak
spectrum as appropriate. In this way, the quasar PSF contribution
could be determined for each annulus analysed and in separate
observations if needed.

Fig. \ref{fig:sbprofilefull} shows surface brightness profiles
extracted from the observation and the quasar PSF simulation for the
energy band $0.5-7\keV$.  These profiles were extracted from
concentric annuli, centred on the quasar, and each $1\asec$ wide.  The
blank sky background was subtracted and the profiles were exposure corrected.  Uncertainties from the best-fit quasar spectral model were
evaluated by running additional PSF simulations and are shown as a
shaded region on the residual surface brightness profile.
Fig. \ref{fig:sbprofilefull} (top) shows that the PSF simulation
predicts a higher surface brightness than the observed profile within
a radius of $1\asec$.  This is due to pileup, which suppresses the
flux of the observed profile.  The bright quasar PSF clearly dominates
over the hot atmosphere to a radius of $4\asec$ and then the PSF wings
comprise $>10\%$ of the total flux to a radius of $20\asec$.  However,
even after subtraction of the PSF, the residual profile shows a strong
peak within a radius of $4\asec$ ($18\kpc$) suggesting that the
surface brightness of the hot atmosphere increases by more than an order of
magnitude.  Given the spatial coincidence with the bright quasar, it
is natural to assume this peak is a result of an undersubtraction of
the quasar PSF.  However, this would require an underestimation by
$\sim35\%$ for the energy band $0.5-7\keV$ (increasing $>200\%$ in a
soft-band).

\begin{figure*}
  \begin{minipage}{\textwidth}
    \centering
    \includegraphics[width=0.45\columnwidth]{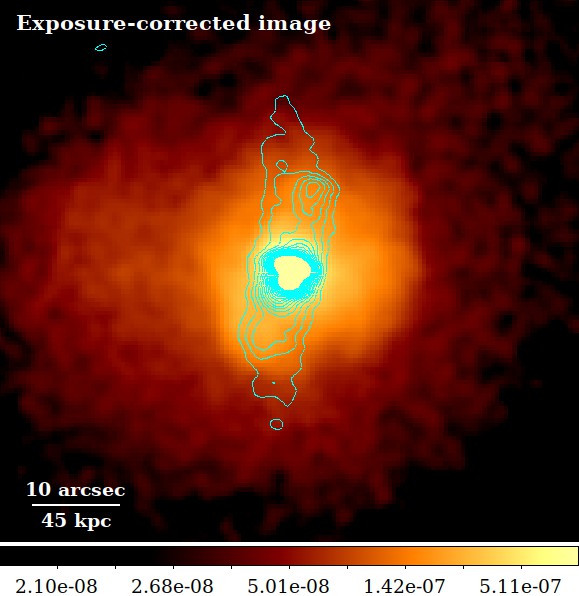}
    \includegraphics[width=0.45\columnwidth]{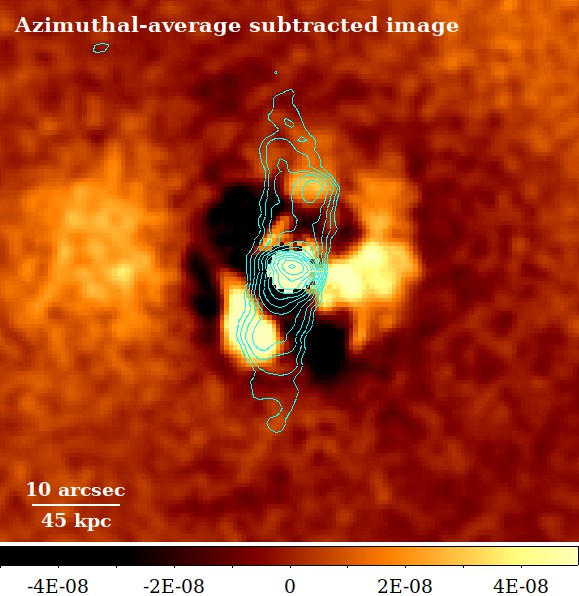}
    \caption{Merged image for the energy range $0.5-7\keV$.  Image is exposure corrected with units of $\expmapcorr$ and smoothed with a 2D Gaussian with $\sigma=1.5\asec$.  Readout streaks were excluded from each separate observation and the range in roll angle ensures that the removal is not readily visible once merged.  VLA $1.4\GHz$ radio contours at $5\sigma,10\sigma,15\sigma ...$ from \citet{Blundell01} are also shown.  N is up and E is to the left.  Right: Residual image where the average surface brightness at each radius, centred on the quasar, has been subtracted.  Note that, whilst this method produces residuals around the quasar, the subtraction using the PSF simulation does not.}
    \label{fig:images}
  \end{minipage}
\end{figure*}


We consider the likelihood of such a significant underestimation of
the quasar PSF.  Whilst this PSF subtraction technique has been
utilized effectively in the presence of bright hot atmospheres in a
number of previous studies (\citealt{Russell10};
\citealt{Siemiginowska10}; \citealt{Bambic23}), the dominance of the
quasar PSF in H1821+643 renders this analysis particularly
challenging.  There are a number of potential systematics to consider.
\textsc{marx} version 5 uses the \textsc{AspectBlur} parameter to blur
the PSF by the uncertainty in the aspect solution.  The observed PSF
has been found to be wider than the equivalent simulated PSF so this
parameter is typically empirically adjusted to match the observation
of interest and is dependent on the source spectrum and the
observation date.  There is also evidence that the soft-band ACIS PSF
($<0.8\keV$) has broadened in observations since $\sim$2017, likely
due to the continued contaminant build-up on the ACIS filters.  The
presence of bright, extended emission makes it impossible to fine-tune
the PSF model to mitigate these problems.  Furthermore, the model
accounting for the change in effective area under the contaminant
build up is continually evolving.  An underestimation of the
contaminant layer would artificially suppress the measured quasar
flux, leading to an undersubtraction of this component and a clear
excess at low energies with a similar slope to the PSF.  We note
however that an underestimation by $\sim200\%$ appears highly unlikely
given \textit{Chandra's} ongoing program of detailed calibration
observations.  We tested the impact of the latest contaminant model
update available in \textsc{CALDB} version 4.10.2 and found that the
measured quasar flux from the readout streak increased by $2.4\%$ for
the energy range $0.5-1\keV$.


The best way to determine the impact of systematic uncertainties on
the PSF simulation is to examine the PSF-subtracted surface
brightness profile in a hard-band.  Whilst the soft-band emission is
often highly structured at the centres of brightest cluster galaxies,
usually due to dense, cool gas filaments and rims around cavities, the
hard-band emission traces the hotter, projected cluster gas, which is
more uniform.  Any substructure on small scales around the quasar in
a hard-band surface brightness profile is therefore likely due to
inaccuracies in the PSF subtraction alone.

Fig. \ref{fig:sbprofiles} (right) shows the surface brightness
profiles extracted from the observation and the quasar PSF simulation
for the energy band $4-7\keV$.  We note that pileup artificially
boosts the observed flux in the hard-band within a radius of $1\asec$
and this is not captured by the PSF simulation.
Fig. \ref{fig:pileupspec} shows how pileup distorts the spectral shape
with piled up soft-band photons detected as an excess in the hard
band.  This region is therefore not considered further.  As discussed
in section \ref{sec:data}, the region from $1-2\asec$ is also likely
affected by a low level of pileup.  This produces the observed spike
in the hard-band surface brightness profile in this region as pileup
boosts the total hard-band flux by $\sim8\%$.  The flux in the
$2-3\asec$ region is not artificially increased by pileup and is
therefore the best indicator of an accurate PSF subtraction.  A modest
drop in the hard-band flux here suggests that the PSF is marginally
oversubtracted by the model.  However, this is not significant given
the uncertainties from the readout streak spectral model (shown by the
grey shaded region).  Beyond the pileup affected regions, the hard
band surface brightness profile has a smooth gradient with no
significant substructure in the region dominated by the quasar
PSF.

Whilst the PSF model is clearly a good match in the hard-band,
the ACIS PSF has broadened significantly at low energies since
$\sim$2017 and this systematic requires a separate evaluation.  Using
the readout streak, which is unaffected by pileup, we extract the
surface brightness for the energy range $0.5-1\keV$ within a radius of
$1\asec$.  The measured surface brightness matches the PSF simulations
in this region within the uncertainties.  Any significant broadening
of the PSF ($\sim0.1\asec$), due to the contaminant buildup or the
\textsc{AspectBlur} parameter, would reduce the measured surface
brightness in this region (spreading the flux to larger radii) and
produce a mismatch with the readout streak result.  These effects are
therefore too small to explain the extended soft band peak.

Fig. \ref{fig:sbprofiles} (left) shows the soft-band surface
brightness profiles.  There is a strong central peak in the residual
profile, which is due to a factor of $\sim30$ increase in the surface
brightness over a few arcsec ($\sim13\kpc$).  The PSF model flux would
need to be increased by a factor of $\sim3$ to remove this peak.  Such a
large increase would be inconsistent with the observed level of pileup
(section \ref{sec:data}) and would still leave a residual soft-band
peak because the quasar PSF is significantly narrower than the soft-band
peak (shown by comparison with the dashed line in
Fig. \ref{fig:sbprofiles} left).

In summary, the soft-band surface brightness of the galaxy's hot
atmosphere increases by a factor of $\sim30$ within the central few
arcsec ($\sim13\kpc$).  This peak cannot be explained by an
undersubtraction of the quasar PSF.  The hard-band surface brightness
profile shows no significant residuals on these scales (excluding
regions affected by pileup).  The soft-band PSF flux would have to be
underestimated by a factor of $\sim3$ and the profile must broadened
by $\sim0.5\asec$ to explain this peak.  These are both inconsistent
with the flux measured from the readout streak spectrum, which is
unaffected by pileup.  The conclusion is a clear, extended soft-band
peak.

Rather than an undersubtraction of the quasar emission, it is more
likely that our PSF model oversubtracts emission.  The strength of the
extended soft-band peak indicates that the readout streak spectrum
will contain a non-negligible contribution from the hot atmosphere.
It was not possible to spectrally separate this from the quasar
emission.  Therefore, the PSF model contains a low level of soft
extended emission that is then effectively subtracted from the
neighbouring annuli.  We consider the impact of this and mitigation in
section \ref{sec:spec}.

\section{Image analysis and surface brightness profiles}
\label{sec:image}



Fig. \ref{fig:images} shows the merged image for the new
\textit{Chandra} dataset and a residual image, where the average
surface brightness at each radius has been subtracted from the image.
The quasar emission dominates the central few arcsec and the extended
emission dominates beyond a radius of $4\asec$.  The extended cluster
atmosphere is clearly asymmetrical about the quasar with bright spurs
of emission along the N-S axis of the radio lobes and to the W to a
radius of $\sim15\asec$.  VLA $1.4\GHz$ contours from
\citet{Blundell01} show that the SE spur is a bright rim around the S
radio lobe and is likely associated with an X-ray surface brightness
depression or cavity.  The N extension similarly extends around the N
radio lobe and is also coincident with an X-ray surface brightness
depression.  Whilst there are other depressions visible in the
residual X-ray image, these appear to be a result of the azimuthal
averaging with clear excess emission at the same radius.  The
cavities are each located $27\kpc$ from the nucleus and have radii of
$\sim14\kpc$.  The W spur appears to terminate at a surface brightness
edge that runs W to S around the cluster core.  This edge is sharpest
to the NW and SE.  Beyond the immediate environment of the radio
lobes, the cluster emission on $100\kpc$ scales is asymmetric, with
greater extension to the E (Fig. \ref{fig:images} right).

\begin{figure}
\centering
\includegraphics[width=0.98\columnwidth]{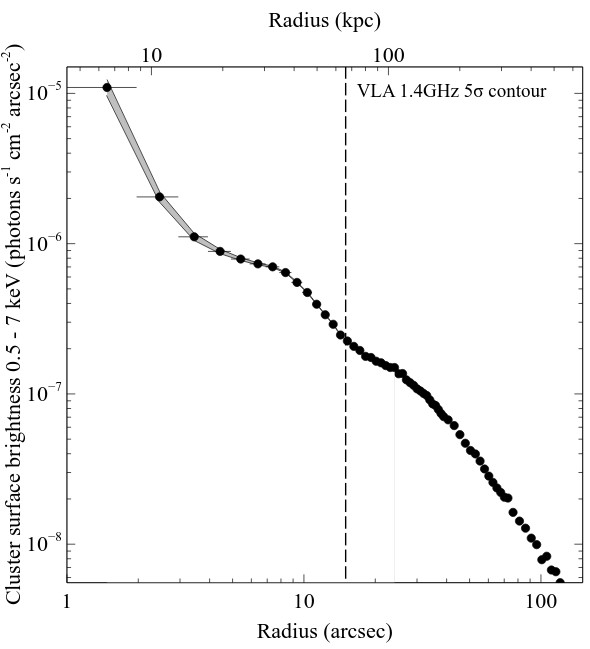}
\caption{Residual surface brightness profile ($\surbri$) in the energy range $0.5-7\keV$ extending to large radius.  The quasar PSF emission has been subtracted.  The approximate extent of the radio emission to the N and S of the quasar is shown by the dashed line (see Fig. \ref{fig:images}).  Uncertainties from the best-fit quasar spectral model are shown by the shaded band.}
\label{fig:sbprofilelgrad}
\end{figure}

The strong peak in the extended emission within the central few arcsec
is only readily apparent once the quasar PSF has been subtracted.
Fig. \ref{fig:sbprofilelgrad} shows the PSF-subtracted surface
brightness profile with an increase of more than an order of magnitude within a
radius of $3\asec$ ($13\kpc$).  This increase is unlikely to be due to the quasar PSF wings (see section \ref{sec:PSFsim}) or scattering of quasar emission off free electrons and dust within the host galaxy.  To demonstrate the latter, we compare the ratio of the ``scattered'' emission, $F_{\mathrm{scatt}}$, to the direct emission from the AGN, $F_{\mathrm{AGN}}$.

\begin{figure}
  \centering
  \includegraphics[width=0.8\columnwidth]{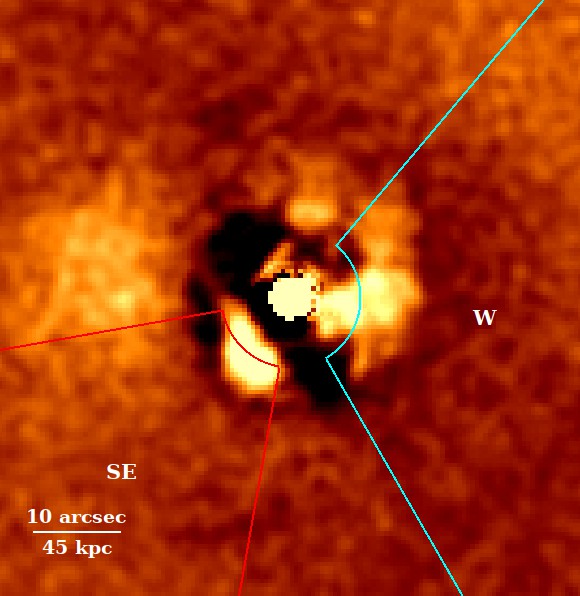}
  \caption{Residual image (see Fig. \ref{fig:images}) with sectors overlaid that cover the surface brightness edge that runs W to S around the cluster core.}
  \label{fig:imgsectors}
\end{figure}

\begin{figure}
  \centering
  \includegraphics[width=0.95\columnwidth]{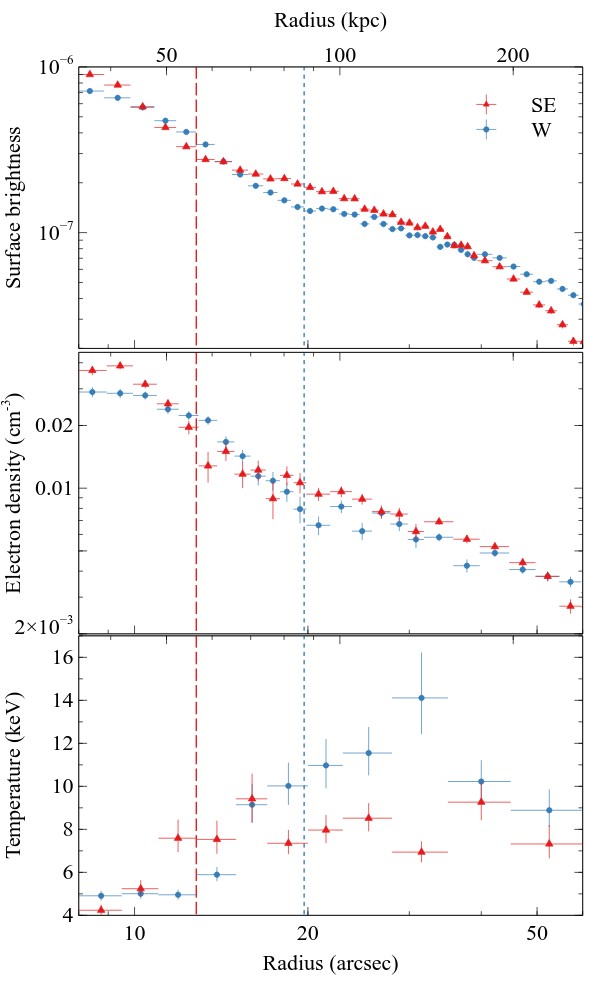}
  \caption{Upper: Surface brightness profiles for the SE and W sectors (see Fig. \ref{fig:imgsectors}) with units of $\surbri$.  Middle: Deprojected electron density profiles.  Lower: Projected gas temperature profiles.  Note that the profiles do not extend into the region affected by the quasar PSF.  The location of surface brightness edges in the SE and W profiles are shown by the vertical lines (red dashed and blue dotted, respectively).}
  \label{fig:sectorprofiles}
\end{figure}

\begin{equation}
  \frac{F_{\mathrm{scatt}}}{F_{\mathrm{AGN}}}=\frac{1}{L_{\mathrm{AGN}}}\int{\left(\frac{L_{\mathrm{AGN}}}{4\pi{r}^{2}}\right)\sigma_{\mathrm{T}}n_{\mathrm{e}}dV},
  \label{eq:fscatt}
\end{equation}

\noindent where $L_{\mathrm{AGN}}$ is the AGN luminosity and $\sigma_{\mathrm{T}}$ is the Thomson scattering cross section.  Assuming a powerlaw form for the electron density profile, 

\begin{equation}
n_{\mathrm{e}}=n_{\mathrm{e,r_{max}}}\left(\frac{r}{r_{\mathrm{max}}}\right)^{-\zeta},
\end{equation}

\noindent we can integrate equation \ref{eq:fscatt} over the radial range that covers the strong increase in observed surface brightness from $r_{\mathrm{min}}=1\asec$ ($4.4\kpc$) to $r_{\mathrm{max}}=4\asec$ ($17.6\kpc$).  This produces an equation of the form

\begin{equation}
\frac{F_{\mathrm{scatt}}}{F_{\mathrm{AGN}}}=\sigma_{\mathrm{T}}n_{\mathrm{e,r_{max}}}\left(\frac{r_{\mathrm{max}}}{1-\zeta}\right)\left[1-\left(\frac{r_{\mathrm{min}}}{r_{\mathrm{max}}}\right)^{1-\zeta}\right].
\end{equation}

\noindent For $n_{\mathrm{e,r_{max}}}\sim0.046\pm0.006\pcmcu$ and $\zeta=3.04\pm0.11$
(appropriate from $1-4\asec$ radius, see section \ref{sec:spec}), we
predict $F_{\mathrm{scatt}}/F_{\mathrm{AGN}}=0.013\pm0.002$.  The observed ratio is nearly an order of magnitude higher, $F_{\mathrm{scatt}}/F_{\mathrm{AGN}}=0.090\pm0.002$.  Therefore, the observed increase in surface brightness is far too bright to be scattered quasar light.  In section \ref{sec:spec}, we show that the strong surface brightness peak is instead from thermal gas emission.

Beyond the quasar's immediate environment, the radio lobes are contained within a
broad surface brightness edge at $\sim15\asec$ ($65\kpc$).  The
filaments and asymmetrical structure can be seen from surface
brightness profiles extracted in sectors (Fig. \ref{fig:imgsectors}).  The selected
sectors demonstrate the shift in radius of the surface brightness edge
from $13\asec$ ($57\kpc$) in the SE sector to $20\asec$ ($88\kpc$) in
the W sector (Fig. \ref{fig:sectorprofiles}).

\section{Spectral analysis}

We mapped the cluster gas properties by extracting spectra in full
annuli, sectors and free-form regions.  The latter covered pixels of similar
surface brightness and were generated with a contour binning
algorithm (\citealt{Sanders06}).  The contribution from the quasar PSF
was modelled with an absorbed powerlaw component.  The powerlaw parameters
were determined by fitting this model to spectra extracted in
identical spatial regions in the PSF simulation.  Each region selected
had at least 5000 counts spread across 22 separate spectra (one from each observation).
Appropriate response files were generated and background spectra were
extracted from the tailored blank sky fields (section \ref{sec:data}).
Spectra from all observations were fit simultaneously in
\textsc{xspec} over the energy range
$0.5-7\keV$ and grouped to ensure a minimum of 20 counts per spectral
channel.  The $\chi^2$-statistic was used to determine the best-fit model.

The cluster emission was modelled with an absorbed \textsc{apec} model
(\citealt{Smith01}).  The absorption of this component was fixed to
the Galactic value (section \ref{sec:quasarspec}) and the redshift was
fixed to $z=0.296$.  The temperature, metallicity and normalization
parameters were left free.  Uncertainties from the best-fit quasar
spectral model were also assessed (section \ref{sec:quasarspec}).
Statistical uncertainties in the PSF model parameters were found to be
less than the statistical uncertainties in the cluster emission
parameters at each radius.

\begin{figure}
  \centering
  \includegraphics[width=0.98\columnwidth]{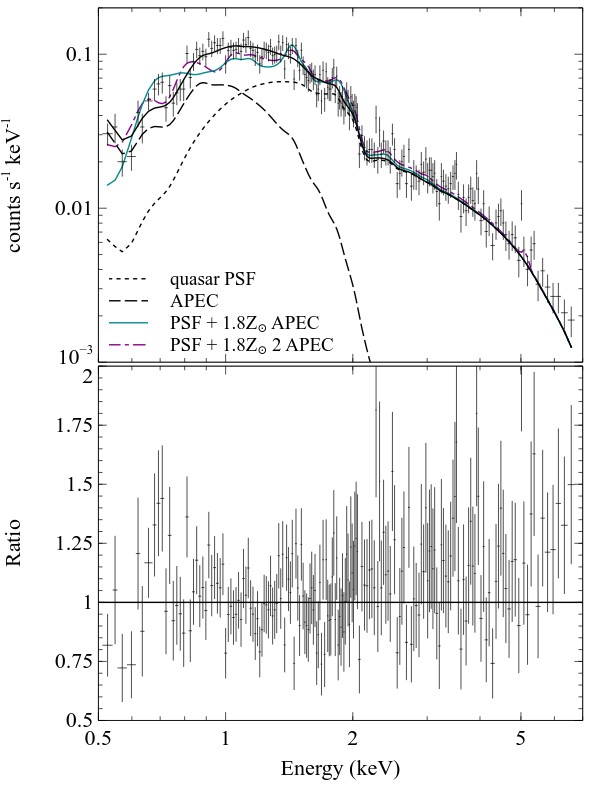}
  \caption{Upper: Spectrum for Obs. ID 21558 extracted from an annulus
    from $1$ to $2\asec$ radius.  The spectrum is grouped with a
    minimum of 20 counts per channel.  The contribution from the
    quasar PSF wings has been determined from the PSF simulation and
    is shown by the dotted line.  The remaining emission is likely
    thermal emission from the ISM and is modelled with an
    \textsc{apec} component (dashed line).  The total model is shown
    by the solid black line.  The best-fit to this single spectrum gives
    $\chi^{2}=190$ for 180 degrees of freedom ($\chi_{\nu}^{2}=1.06$).  The total models for fixed metallicity of $1.8\Zsun$ and one ($\chi_{\nu}^{2}=1.81$) and two ($\chi_{\nu}^{2}=1.51$) \textsc{apec} components are shown by the solid cyan and dash-dotted magenta lines, respectively.  Lower: ratio of the data to the best-fit model.  The strongest
    residuals are above $5\keV$ and therefore most likely due to
    pileup as expected for this region (see section \ref{sec:data}).}
  \label{fig:egspec}
\end{figure}

All best-fit parameters were determined by fitting simultaneously to
spectra from the 22 separate observations.  The continual reduction in
effective area due to the contaminant build up prevents us from adding
spectra together.  However, in Fig. \ref{fig:egspec}, we show a
spectrum extracted from $1-2\asec$ radius from a long individual
observation with a typical PSF flux (Obs. ID 21558).  This spectrum
has over 7000 counts and was grouped with a minimum of 20 counts per
channel for clarity.  We overplot the corresponding best-fit model and
individual components to demonstrate the quality of the fit and the
impact of the quasar PSF.  The best fit parameters are all consistent
with the results for the simultaneous fit to all spectra.  The
best-fit to the single spectrum gives $\chi^{2}=190$ for 180 degrees
of freedom ($\chi_{\nu}^{2}=1.06$).  The strongest residuals are above
$5\keV$ and therefore most likely due to pileup (see section
\ref{sec:data}).  Modest residuals at $\sim0.7\keV$ are also visible
but the single \textsc{apec} model appears to be a reasonable fit to
the soft emission.

\begin{figure}
  \centering
  \includegraphics[width=0.98\columnwidth]{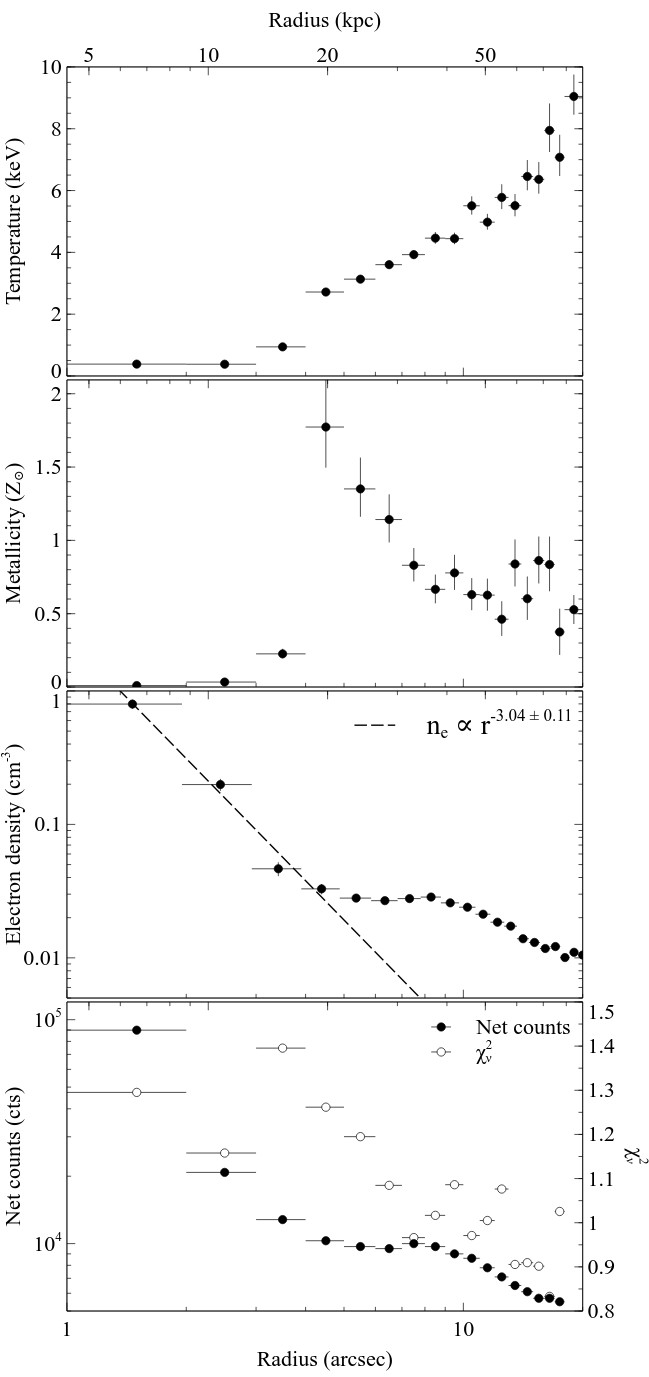}
  \caption{Upper panels: projected temperature, projected metallicity and deprojected electron density profiles for the cluster gas.  Bottom panel: net counts and $\chi_{\nu}^2$ for each annulus.  The quasar PSF contribution has been subtracted from each plot.}
  \label{fig:profiles}
\end{figure}

\subsection{Radial profiles}
\label{sec:spec}

Fig. \ref{fig:profiles} shows the projected temperature, projected
metallicity and deprojected electron density profiles for the cluster
gas.  The deprojected electron density profile was produced using the
\textsc{dsdeproj} routine, which assumes spherical symmetry to subtract
the projected contribution from each successive annulus
(\citealt{Sanders07}; \citealt{Russell08}).  Regardless of the excess
hard photons due to pileup ($1-2\asec$) or the modest oversubtraction
of the quasar PSF ($2-3\asec$), the gas temperature drops precipitously to $0.4\keV$
for both annuli inside $3\asec$ ($13\kpc$).  This is very close to the
lower bound of the \textsc{apec} model and therefore in effect is an
upper limit on the gas temperature.  From $3$ to $4\asec$, the gas
temperature increases to $0.94\pm0.03\keV$.  Then, at $4\asec$ radius
($18\kpc$), the temperature sharply increases by a factor of 3 to
$2.72^{+0.10}_{-0.09}\keV$.  Beyond this radius, the temperature
increase is smoother and extends up to $9.0^{+0.7}_{-0.6}\keV$ at a
radius of $20\asec$ ($90\kpc$).


The metallicity profile at the centre of H1821+643 is similarly
striking.  The metallicity within $3\asec$ ($13\kpc$) of the quasar is
essentially an upper limit of $0.04\Zsun$.  There is a modest increase
from $3$ to $4\asec$ to an abundance of $0.23\pm0.03\Zsun$.  The
metallicity then increases by a factor of roughly 10 to a peak of
$1.8\pm0.3\Zsun$ and then decreases with radius to the average value
of $0.4\Zsun$ at a radius approaching $100\kpc$.  If we fix the metallicity in the innermost radial bin to $1.8\Zsun$, the corresponding best-fit model has $\chi_{\nu}^2=1.95$ ($\chi^2=5689$
for 2916 degrees of freedom) compared to $\chi_{\nu}^2=1.30$
($\chi^2=3775$ for 2915 degrees of freedom) for a free, low
metallicity value.  Similarly, even a modest metallicity of $0.4\Zsun$ is excluded.  The
corresponding best-fit model has $\chi_{\nu}^2=1.73$ ($\chi^2=5046$
for 2916 degrees of freedom).  Additional \textsc{apec} or \textsc{mkcflow}
components, or a free Galactic absorption component, did not
significantly improve the fit or alter the low metallicity value.  The
only exception is for $3-4\asec$ where an additional \textsc{apec}
component at $\sim0.4\keV$ does produce a better fit with
$\chi_{\nu}^2=1.06$ but the metallicity remains low.

Although low metallicity would result from fitting a thermal plasma
model to quasar continuum emission, we have demonstrated in section
\ref{sec:PSFsim} that our model accounts for the vast majority of the
quasar emission.  Another possible systematic is the low level
component of soft extended emission in the PSF model, which is then
effectively oversubtracted from the surrounding annuli (see section
\ref{sec:PSFsim}).  For gas temperatures below $\sim2\keV$, the blend
of Fe L lines near $1\keV$ (redshifted to $0.75\keV$) is particularly
prominent and may be affected by this oversubtraction.  We therefore
repeated the spectral fitting analysis but excluded emission below
$1\keV$ (see e.g. \citealt{Ghizzardi21}).  We obtain similarly tight
constraints on the metallicity in the innermost two annuli and
similarly low temperature values.  Fig. \ref{fig:egspec} shows that
these tight constraints are due to the absence of a clear Mg line at
$\sim1.1\keV$ (rest frame $1.4\keV$) and Si and S lines at
$1.5-2.3\keV$ (rest frame $2-3\keV$).  The low metallicity likely has
a physical origin, which is confirmed by Suzaku observations of
H1821+643 that also indicate subsolar metallicity
(\citealt{Reynolds14}).

The hot atmosphere within a radius of $4\asec$ ($18\kpc$) has likely
been photo-ionized by the quasar emission.  The ionization parameter
$\xi=L/n_{\mathrm{e}}r^2$ where $L$ is the quasar luminosity above
$15\eV$ ($1.1\pm0.1\times10^{46}\ergps$) and $n_{\mathrm{e}}$ is the
electron density at a radius of $r$.  At a radius of $20\kpc$,
$\xi=90\pm10\ionpar$.  Using \textsc{xstar}, \citet{Kallman01} show that, at
$\xi=90\pm10\ionpar$, Mg, O, Si and S ions will be almost completely
photo-ionized and Fe will be partly photo-ionized.  If the quasar
emission is beamed out of the line of sight, then we expect strong
asymmetries with ions in the path of the beam completely
photo-ionized.  \citet{Russell10} also used Cloudy simulations (\citealt{Chatzikos23}) to
demonstrate that the quasar emission from H1821+643 will produce a
significant amount of photoionization in this region and the line
emission will be suppressed.  The previous
\textit{Chandra} observations could not distinguish this effect from
the Fe bias created by fitting multi-temperature gas with a single
temperature model.  With the new dataset, the dramatic decline in
metallicity at small radii is revealed.  This cannot be explained by
Fe bias in multi-temperature gas and is instead due to photo-ionization.


The deprojected electron density increases by a factor of $\sim25$
from a radius of $20$ to $6\kpc$ ($5$ to $1.5\asec$) at the centre of
H1821+643 (Fig. \ref{fig:profiles}).  We fit a powerlaw model to this
section of the density profile and find a best-fit where
$n_{\mathrm{e}}=240^{+70}_{-60}\left(r/\mathrm{kpc}\right)^{-3.04\pm0.11}\pcmcu$.
If the gas temperature has dropped to $0.4\keV$ in this region, as
indicated by the temperature profile, we would expect a steep rise in
density as the gas is compressed by the outer atmosphere.  For
$0.4\keV$ gas in an isothermal potential with galaxy line of sight
velocity $300\kmps$ (based on the total galaxy mass
estimates in \citealt{Fukuchi22}), the gas density in a hydrostatic
atmosphere should vary as
$r^{-2{\mu}m_{\mathrm{H}}\sigma^{2}/k_{\mathrm{B}}T}$,
where the exponent is $-2.9$.  This is consistent with the gradient of
the observed density profile (Fig. \ref{fig:profiles}).  The observed rapid increase in gas
density and decrease in temperature therefore appear consistent and
suggest that the emission in this region is thermal ISM.

\begin{figure}
  \centering
  \includegraphics[width=0.98\columnwidth]{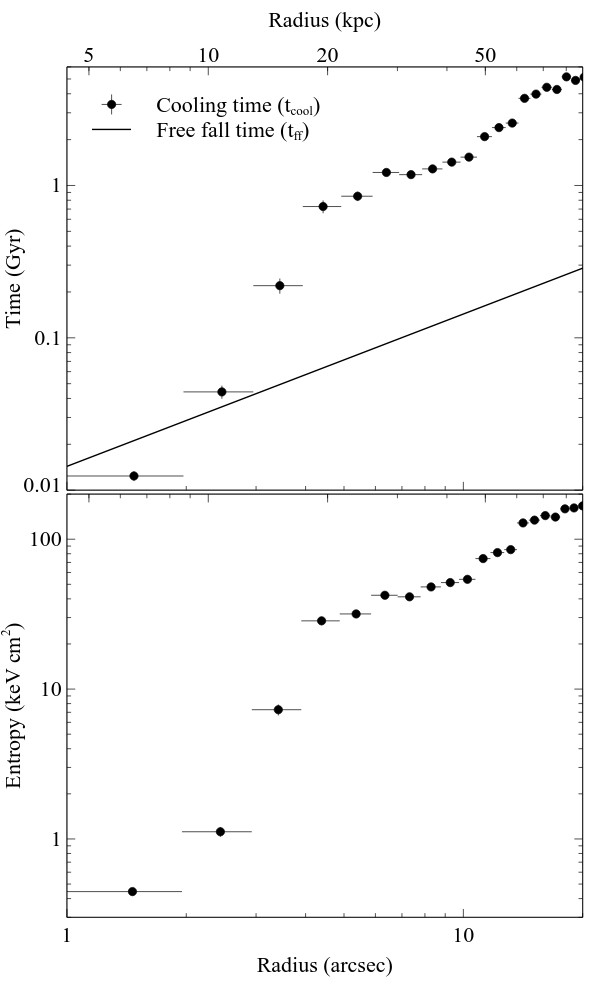}
  \caption{Upper: Radiative cooling time and free fall time profiles.  Lower: Entropy profile.}
  \label{fig:tcool}
\end{figure}

Fig. \ref{fig:tcool} shows the radiative cooling time and free fall time profiles.  The radiative cooling time,

\begin{equation}
t_{\mathrm{cool}}=\frac{3}{2}\frac{nk_{\mathrm{B}}T}{n_{\mathrm{e}}n_{\mathrm{H}}\Lambda\left(T,Z\right)},
\end{equation}

\noindent where $n$ is the gas density, $n_{\mathrm{e}}$ is the
electron density, $T$ is the temperature, $n_{\mathrm{H}}$ is the ion
density and $\Lambda(T,Z)$ is the cooling function, which depends on
the gas temperature and metallicity.  If Fe is not fully
photo-ionized within a radius of $18\kpc$, the cooling function will
be underestimated here by up to an order of magnitude (for
$1.8\Zsun$).  The free fall time profile was calculated assuming an
isothermal potential and $\sigma=300\kmps$ (see
e.g. \citealt{Hogan17}).  The cooling time drops sharply below a Gyr
within the central $4\asec$ ($18\kpc$) and falls to $12\pm1\Myr$ at
$1.5\asec$ ($6.5\kpc$).  The cooling time falls below the free fall
time here.  The centre of H1821+643 is therefore undergoing runaway or
catastrophic cooling.  If the cooling function has been
underestimated, the cooling time within a radius of $18\kpc$ is likely
shorter by up to an order of magnitude and will fall below the free
fall time at this larger radius.

Fig. \ref{fig:tcool} also shows the cluster entropy profile calculated from
the deprojected electron density and projected temperature profiles as
$K=T/n_{\mathrm{e}}^{2/3}$.  The innermost value drops to
$0.44\pm0.02\keVcmsq$ at a radius of $6.5\kpc$.  This is far below the
typical values for cool core clusters at similar radii of
$\sim10\keVcmsq$ and even below the previous record of $\sim2\keVcmsq$
found at this radius in the Phoenix cluster (\citealt{McDonald19}).

\subsection{Contour binning maps}
\label{sec:contour}

Beyond the quasar's immediate environment, we map the gas properties
using a contour binning algorithm (\citealt{Sanders06}), which
generates spatial regions by grouping together neighbouring pixels
with similar surface brightness.  Pixels were added until each region
contained at least 1000 counts ($S/N=32$).  The quasar and its immediate environment to a radius of $4\asec$ was excluded.  This region can only be analysed with full annuli because of PSF asymmetries (see section \ref{sec:data}).  Each spectrum was fitted
with an absorbed \textsc{apec} model as described above.  The
uncertainties on the temperature value are typically $10\%$ at
$3-4\keV$, $20-30\%$ at $6-8\keV$ and $40-50\%$ at $10-15\keV$.  The
uncertainties on the normalization are typically $5-10\%$.
Pseudo-pressure was generated by multiplying the temperature and
square root of the normalization.  Whilst
the metallicity was left as a free parameter, the corresponding map
does not reveal significant structure beyond the radial variation
shown in Fig. \ref{fig:profiles}.

\begin{figure}
  \centering
  \includegraphics[width=0.98\columnwidth]{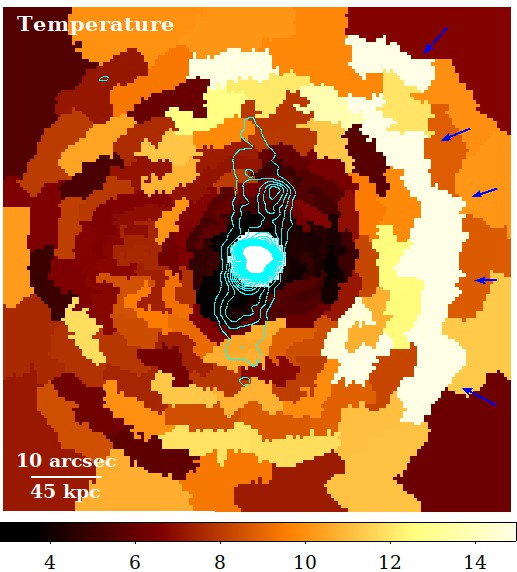}
  \includegraphics[width=0.98\columnwidth]{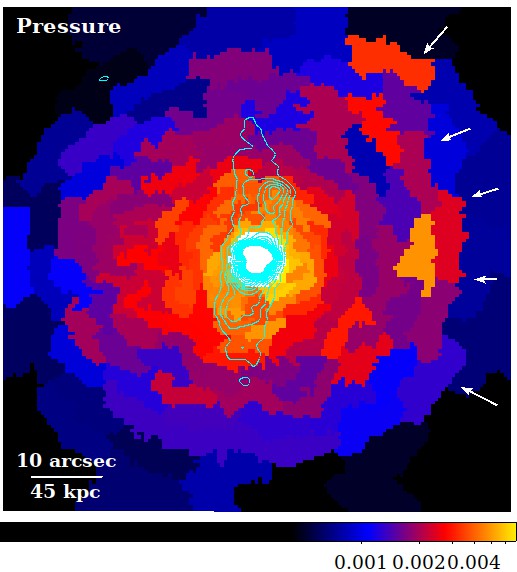}
  \caption{Upper: Projected temperature map (keV).  Lower: Pseudo-pressure map ($\pseudoP$).  The quasar and its immediate environment to a radius of $4\asec$ have been excluded (central white circle).  The radio contours from Fig. \ref{fig:images} are shown overlaid and a hot, relatively high pressure region to the W and NW is marked by the arrows.}
  \label{fig:maps}
\end{figure}

Fig. \ref{fig:maps} shows the resulting temperature and pressure maps.
The low-temperature gas, $3-4\keV$, at the centre clearly extends
along the radio jet axis, particularly along the SE edge of the S
radio lobe.  The temperature then increases sharply to $7-9\keV$ at a
radius of $12\asec$.  This increase coincides with the surface
brightness edge detected in the SE sector.  This structure is likely to be a bright, cool rim of gas around a radio bubble (Fig. \ref{fig:sectorprofiles}).  The bright rim above the N radio lobe has a similarly low temperature.  
A further extension of
cool gas is visible to the W of the quasar to a radius of
$\sim18\asec$ ($\sim80\kpc$).  At this radius, the temperature then
increases sharply from $4-5\keV$ to $>10\keV$.
Fig. \ref{fig:sectorprofiles} shows a surface brightness edge at this
radius, therefore this structure is a cold front.  The
temperature map also shows a hot region at $>10\keV$ that forms an arc
beyond the W cold front.  Although there is no clear surface
brightness edge at this radius in the W sector, the temperature
increase is significant and suggests that this may be a shock-heated
region.  \citet{Bonafede14} proposed that H1821+643 has undergone an
off-axis or minor merger because it hosts a giant radio halo.  Based
on optical imaging, \citet{HutchingsNeff91} suggested H1821+643 was in
the late stages of a mild tidal event or merger.  The W cold front,
which appears unrelated to the radio lobes, the E extension of the
cluster emission on large scales and the shock-heated region to the W
are consistent with this picture.

\section{Discussion}
\label{sec:disc}

We have presented a new $582\ks$ short frame-time \textit{Chandra} observation
of the quasar H1821+643, which lies at the centre of a strong cool
core cluster.  By carefully modelling the quasar PSF, we show that the
surface brightness of the central galaxy's hot atmosphere increases by
a factor of $\sim30$ within the central few arcsec ($\sim13\kpc$) and
the gas temperature drops rapidly to $<0.4\keV$.  The radiative
cooling time of the gas is only $12\pm1\Myr$ at a radius of $6.5\kpc$
and falls below the free fall time.  We have demonstrated that this clear
transition in the gas properties cannot be due to an undersubtraction of the quasar PSF or
scattered quasar light and is instead consistent with runaway cooling
of the galaxy's hot atmosphere.

\subsection{Runaway cooling of the hot atmosphere}

The multi-wavelength data for H1821+643 are also consistent with rapid
cooling of the hot gas.  With a total IR luminosity of
$L_{\mathrm{IR,tot}}=4.6\times10^{46}\ergps=1.1\times10^{13}\Lsun$
(\citealt{Ichikawa19}), the central galaxy is categorized as a
hyperluminous IR galaxy.  H1821+643 is likely in a strong starburst
phase with an estimated star formation rate of
$120^{+120}_{-60}\Msunpyr$ from \textit{HST} [O~\textsc{ii}] maps with
the quasar emission subtracted (\citealt{Calzadilla22}).  Alternative
estimates from polycyclic aromatic hydrocarbon (PAH) $7.7\mum$
emissions (\citealt{Ruiz13}) and SED fitting studies
(\citealt{Farrah02,Fukuchi22}) confirm a high rate of star
formation.  This places H1821+643 amongst the most strongly
star-forming brightest cluster galaxies, which include Abell\,1835,
IRAS\,09104+4109 and MACS\,1931.8+2634.  Only the Phoenix cluster, at
$600^{+300}_{-200}\Msunpyr$, exceeds this group (\citealt{McDonald12,Calzadilla22}).

CARMA observations detect strong CO(1-0) emission from
$\left(8.0\pm1.7\right)\times10^9\Msun$ of molecular gas extending up
to $\sim4\asec$ ($18\kpc$) to the SE of the quasar
(\citealt{Aravena11}, assuming an $X_{\mathrm{CO}}$ factor that is
typical of local luminous IR galaxies).  Although it was not possible
to resolve asymmetries in the hot gas emission on these scales, we
note that the molecular gas aligns with the SE radio lobe and
extension of cool X-ray gas around the bubble rim.  This is indicative
of stimulated cooling where the radio bubbles promote gas cooling by
lifting low-entropy gas (\citealt{McNamara16}).  Given uncertainties
in the CO-to-H$_2$ conversion factor and CO emission line ratios, the
molecular gas mass in H1821+643 is comparable to the most gas-rich
central galaxies, including the Phoenix cluster
($2.1\pm0.3\times10^{10}\Msun$, \citealt{Russell17}) and A1835
($9.9\pm0.5\times10^{9}\Msun$ updated for a LIRG $X_{\mathrm{CO}}$ factor,
\citealt{McNamara14}).

Optical and near-IR observations of the central galaxy are strongly
impacted by the quasar emission but appear to support this picture.
\citet{Floyd04} analyse \textit{HST} WFPC2 observations of H1821+643
using the F791W filter, which covers the bright H$\alpha$ emission
line (see e.g. long slit spectrum from \citealt{Kolman93}).
\citet{Floyd04} detect a bright structure extending $4\asec$ to the SE
of the nucleus.  Although they note that this may be a PSF artefact,
the spatial coincidence with the molecular gas emission suggests this
may instead be ionized gas emission from the surface layers of a
cool gas nebula.  Herschel/PACS observations of H1821+643 found
extended [O~\textsc{i}] emission (tens of kpc across), which is also
likely from cool gas (\citealt{Fernandez16}).


For runaway cooling, the gas flows steadily inwards as it cools,
thereby maintaining pressure support of the overlying layers
(e.g. \citealt{Fabian77,Nulsen86}).  Assuming a steady-state, spherical inflow
of gas, we can use the gas density and temperature profiles at the
centre of H1821+643 to determine the radial flow speed and mass inflow
rate.  In terms of entropy, $S$, the energy equation for this gas flow
is

\begin{equation}
\rho T v_{\mathrm{r}} \left(\frac{\mathrm{d}S}{\mathrm{d}r}\right) = n_{\mathrm{e}} n_{\mathrm{H}} \Lambda,
\end{equation}

\noindent where $\rho$ is the gas density, $T$ is the gas temperature, $v_{\mathrm{r}}$ is the inflow speed and $n_{\mathrm{e}} n_{\mathrm{H}}\Lambda$ is the energy radiated.  We can rewrite this expression in terms of gas cooling time, $t_{\mathrm{cool}}$, and cluster entropy $K=k_{\mathrm{B}}T/n_{\mathrm{e}}^{2/3}$ (e.g. \citealt{Voit05}) as

\begin{equation}
v_{\mathrm{r}} \left(\frac{\mathrm{d\ln}K}{\mathrm{d\ln}r}\right) = \frac{r}{t_{\mathrm{cool}}}.
\end{equation}

\noindent At a radius of $\sim8\kpc$, where $t_{\mathrm{cool}}\sim
t_{\mathrm{ff}}$, the entropy gradient
$\mathrm{d\ln}K/\mathrm{d\ln}r=1.8\pm0.2$ and the cooling time
$t_{\mathrm{cool}}=12\pm1\Myr$.  The inflow velocity is then
$360\pm50\kmps$.  This is roughly equal to the sound speed of
$320\pm20\kmps$ for gas at $0.4\keV$.  The sonic radius is therefore
resolved in H1821+643, assuming that the flow is steady and spherical.
As discussed in section \ref{sec:spec}, if Fe is only
  partially photo-ionized within a radius of $18\kpc$, the cooling
  function, $\lambda$, will be underestimated by up to an order of magnitude and
  the cooling time will be up to an order of magnitude shorter.  The
  inflow velocity and sonic radius are therefore effectively lower
  limits.

Mass conservation for the gas inflow is given by

\begin{equation}
\label{eq:masscons}
\dot{m}_{\mathrm{inflow}}=4\pi \rho v_{\mathrm{r}}r^2.
\end{equation}

\noindent The spherical mass inflow rate at a radius of $8\kpc$ is
then $3500\pm500\Msunpyr$.  In addition to systematic uncertainties
from the metallicity and the PSF model (sections \ref{sec:spec} and
\ref{sec:PSFsim}, respectively), the gas flow is unlikely to be
completely steady and spherical given the interactions with the radio
lobes.  The mass inflow rate is also dependent on the cooling function.  We therefore compare the mass inflow rate with the classical mass deposition rate, which instead relies on conservation of energy and may be more appropriate.  The classical mass deposition rate

\begin{equation}
\dot{M}=\frac{2}{5}\frac{\mu{m_{\mathrm{H}}}L\left(r<r_{\mathrm{cool}}\right)}{k_{\mathrm{B}}T\left(r_{\mathrm{cool}}\right)},
\end{equation}

\noindent where $r_{\mathrm{cool}}$ is the cooling radius,
  $L(r<r_{\mathrm{cool}})$ is the gas luminosity within
  $r_{\mathrm{cool}}$ and $T(r_{\mathrm{cool}})$ is the gas temperature
  at $r_{\mathrm{cool}}$.  For $r_{\mathrm{cool}}=20\kpc$,
  $\dot{M}=3700\pm200\Msunpyr$.  This is consistent with the mass inflow rate and suggests only a modest underestimation of the cooling function.

Both calculations of the mass flow rate are overestimates because they
discount the contribution of the gravitational potential energy to the
power radiated.  We therefore treat these values as order of magnitude
estimates.  Nevertheless, the mass flow rate can easily supply the
observed star formation rate of $120^{+120}_{-60}\Msunpyr$ in
H1821+643 and the required accretion rate onto the SMBH of
$\sim40\Msunpyr$ (\citealt{Russell10}).

Deep \textit{Suzaku} observations are consistent with this result.
\citet{Reynolds14} study the quasar spectrum with \textit{Suzaku} and
find a soft excess and significantly sub-solar abundances, which they
suggest is due to a Compton-induced cooling flow.  Unfortunately, the
very low metallicity and quasar variability preclude any meaningful
constraint on the gas cooling rate from archival XMM-Newton RGS
observations.

\subsection{Comparison with other strong cool core clusters}

\begin{figure}
\centering
  \includegraphics[width=0.98\columnwidth]{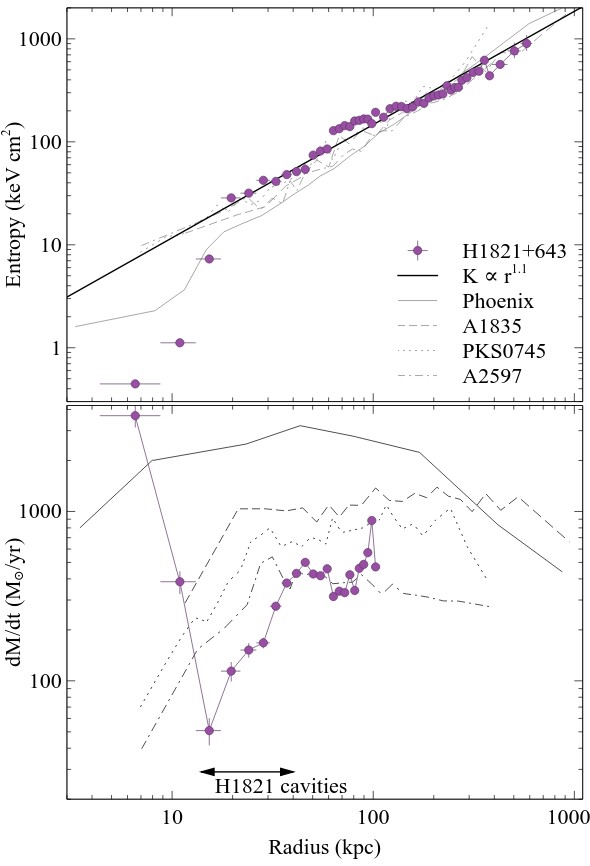}
  \caption{Upper: Cluster entropy profiles for H1821+643 and example strong
    cool core clusters A2597, PKS0745, A1835 and Phoenix.  The
    baseline entropy profile from \citet{Voit05} is shown for
    comparison.  The entropy profile from the previous
    \textit{Chandra} observation of H1821+643 covers the radial range
    $15-600\kpc$ and is used to extend the profile beyond a radius of $100\kpc$
    (\citealt{Russell10,Walker14}).  Lower: Mass flow rate as a function of radius, where the entropy profile for each target was fit with a broken powerlaw model.}
  \label{fig:massdep}
\end{figure}

Fig. \ref{fig:massdep} shows the cluster entropy and mass flow rate
profiles for H1821+643 compared to other strong cool core clusters.
The entropy profile from \citet{Russell10} (see also
\citealt{Walker14}) covers the radial range $15-600\kpc$ and was used
to extend the entropy profile from this analysis beyond $100\kpc$.  We
compare the entropy profiles with the baseline entropy profile,
$K{\propto}r^{1.1}$ from \citet{Voit05}.  The baseline profile is
predicted from numerical simulations of structure formation that
include only gravitational processes (no feedback).  The entropy
profiles for the vast majority of cool core clusters tend to this
baseline at large radii and then diverge to a shallower gradient or
plateau within the central $\sim100\kpc$ where feedback increasingly
dominates (e.g. \citealt{Donahue06,Cavagnolo09}).  Using the previous
\textit{Chandra} analysis of H1821+643, which covered the radial range
$15-600\kpc$, \citet{Walker14} showed that the entropy profile for
H1821+643 falls significantly below this baseline on small scales and
concluded that significant cooling had occurred.  With an entropy
profile that now extends to the central few kpc, we reveal that the
entropy drops by nearly two orders of magnitude within the central
$20\kpc$ and even falls below the previous record of $2\keVcmsq$ from
the other known exception, the Phoenix cluster.

Fig. \ref{fig:massdep} shows the mass flow rate as a function of
radius, where the entropy profile was fitted with a broken powerlaw
model (\citealt{Babyk18}).  At small radii, the vast majority of cool
core clusters show strong suppression of the mass flow rate by
over an order of magnitude as a result of AGN feedback (for example
A2597, PKS0745 and A1835).  The massive Phoenix cluster shows a weak
suppression at small radii by a factor of $\sim5$, which indicates
strong cooling and an underpowered AGN (\citealt{McDonald19}).  In
contrast, the mass deposition rate in H1821+643 increases rapidly to
$>1000\Msunpyr$ within $20\kpc$ and the minimum of $35\pm5\Msunpyr$
occurs at larger radii, coincident with the cavities from $15-33\kpc$.
AGN feedback appears to be heating the hot atmosphere on tens of kpc
scales in H1821+643 but cannot permeate the rapidly cooling region
inside $\sim10\kpc$.

\subsection{Runaway cooling beyond the capabilities of Compton cooling}


Runaway cooling in H1821+643 extends to much larger radii than can be
explained by Compton cooling by the quasar radiation.  Much of the
quasar power is radiated in the UV.  If Compton scattering between UV
photons and hot electrons dominates over bremsstrahlung cooling at the
galaxy centre, this will rapidly cool the hot atmosphere to the Compton temperature $T_{\mathrm{Compton}}\sim0.4\keV$ for H1821+643 (\citealt{Fabian90,Russell10}).  The Compton cooling timescale, $t_{\mathrm{Compton}}$, is given by

\begin{equation}
  t_{\mathrm{Compton}}=\frac{3n}{8n_{\mathrm{e}}}\frac{m_{\mathrm{e}}c^{2}}{\sigma_{\mathrm{T}}F},
  \label{eq:tCompton}
\end{equation}

\noindent where $n$ is the total particle density, $n_{\mathrm{e}}$ is
the electron density, $m_{\mathrm{e}}$ is the electron mass, $c$ is
the speed of light, $\sigma_{\mathrm{T}}$ is the Thomson scattering
cross section and $F$ is the quasar flux.  For Compton cooling to be significant, the Compton cooling timescale must be shorter than the thermal bremsstrahlung radiative cooling timescale,

\begin{equation}
  t_{\mathrm{Compton}}<t_{\mathrm{cool}}.
  \label{eq:comp}
\end{equation}

Combining equations \ref{eq:tCompton} and \ref{eq:comp} gives a lower limit on the required quasar flux,

\begin{equation}
F>\frac{3n}{8n_{\mathrm{e}}}\frac{m_{\mathrm{e}}c^{2}}{\sigma_{\mathrm{T}}t_{\mathrm{cool}}}.
\end{equation}

\noindent By multiplying by $4{\pi}r^2$, this can be expressed as a lower limit on the quasar luminosity,

\begin{equation}
  L>\frac{3n}{8n_{\mathrm{e}}}\frac{m_{\mathrm{e}}c^{2}}{\sigma_{\mathrm{T}}t_{\mathrm{cool}}}4{\pi}r^2
\end{equation}

\noindent where $r$ is the distance from the quasar.  Or equivalently

\begin{equation}
L>3\times10^{45}\left(\frac{r}{\mathrm{kpc}}\right)^2\left(\frac{t_{\mathrm{cool}}}{\mathrm{Gyr}}\right)^{-1} \ergps,
\end{equation}

\noindent For $r=10\kpc$ and $t_{\mathrm{cool}}=50\Myr$ (Fig. \ref{fig:tcool}), the minimum quasar luminosity required to cool this large region is $\sim6\times10^{48}\ergps$.  This is an order of magnitude greater than the measured quasar bolometric luminosity of $3\times10^{47}\ergps$ (\citealt{Fukuchi22}) and rivals the most luminous quasars known (e.g. \citealt{Onken20,Schindler21}).

The quasar may have previously been this luminous but this would be difficult to sustain.  The minimum time required for a significant change in $k_{\mathrm{B}}T-k_{\mathrm{B}}T_{\mathrm{Compton}}$ is $t_{\mathrm{Compton}}$, where $T$ is the temperature of the hot atmosphere.  The total energy that must be radiated by the quasar to cause this significant change in the gas temperature is then

\begin{equation}
Lt_{\mathrm{Compton}}>\frac{3n}{8n_{\mathrm{e}}}\frac{m_{\mathrm{e}}c^{2}}{\sigma_{\mathrm{T}}}4{\pi}r^2.
\end{equation}

\noindent For $r=10\kpc$, $Lt_{\mathrm{Compton}}>10^{64}\erg$.  For an
accretion efficiency of $10\%$, the SMBH would then have to accrete
$6\times10^{10}\Msun$ in $<100\Myr$.  The resulting black hole mass is
close to the limit for luminous accretion given the measured spin
(\citealt{King16,SiskReynes22}) and appears implausible.  The
estimated SMBH mass for H1821+643 is over an order of magnitude lower
at $3.9^{+3.9}_{-2.0}\times10^{9}\Msun$ and was determined from the
broad H$\beta$ emission line and virial, `single-epoch' presciptions
(\citealt{Koss17}).  The Eddington luminosity of this supermassive
black hole is an order of magnitude below the minimum quasar
luminosity required to cool the hot atmosphere to a radius of
$10\kpc$.  Whilst Compton cooling likely enhances cooling on kpc
scales, this mechanism cannot account for rapid gas cooling to a
radius of $10\kpc$.

\subsection{An underpowered AGN?}


Instead of an enhanced cooling explanation, the AGN appears to be
underheating the core.  In section \ref{sec:image}, we identify two cavities to the SE and N of the nucleus
that are spatially coincident with radio lobes at $1.4\GHz$.
Assuming spherical cavities and using the measured thermal pressure of
$\sim3.1\pm0.2\times10^{-10}\ergpcmcu$ at a radius of $27\kpc$, we estimate
that the energy required to inflate each cavity
$E_{\mathrm{cav}}=4pV=\sim3\times10^{59}\erg$.  Using the sound speed
timescale of $27\pm2\Myr$, the total cavity power $P_{\mathrm{cav}}=7\times10^{44}\ergps$.  The uncertainties on the
cavity power are dominated by the cavity volume, including the unknown
extent along the line of sight.  We estimate uncertainties of at least
a factor of 2 in the cavity power.

The cavity power appears low compared to cooling losses of
$2.44\pm0.08\times10^{45}\ergps$ within a radius of $20\kpc$ and a
total of $\sim3.5\times10^{45}\ergps$ within the cooling radius of
$90\kpc$ (the radius at which the cooling time is $7.7\Gyr$ or the
time since $z=1$).  The cavity power appears insufficient to balance
radiative losses by a factor of $\sim5$.  Furthermore, Fig. \ref{fig:massdep}
suggests that the cavity power is dissipated on scales of tens of kpc
and has limited impact on the rapidly cooling region within a radius
of $10\kpc$.


High velocity quasar winds could provide additional mechanical
feedback.  However, no obvious evidence for X-ray absorbing winds was
found in \textit{Chandra} HETG and LETG spectra from 2001 or in FUSE
spectra from 1999 (\citealt{Oegerle00,Fang02,Mathur03}).  Outflows
could be oriented in the plane of the sky and not evident along our
line of sight.  Whilst strong and broad absorption signatures from
winds can be detected in CCD spectra, the variable quasar continuum
emission and evolving contaminant correction made it very difficult to
detect additional absorption above the Galactic column density.

\citet{McDonald18} suggest that the central SMBHs in the most massive
galaxy clusters will be undersized compared to the mass of the cool
core (see also \citealt{Calzadilla22}).  This naturally results from a merger scenario, where low
entropy gas from the smaller halo sinks to the centre of the larger
halo much faster than dynamical friction can merge the central
galaxies.  This may explain the weak suppression of the cooling flow
at the centre of the Phoenix cluster (\citealt{McDonald19}, total cluster mass
$\sim2\sim10^{15}\Msun$).  With a total mass of
$0.6-1\times10^{15}\Msun$ (\citealt{Planck11}; \citealt{Walker14}),
the galaxy cluster hosting H1821+643 is also a particularly massive system and the large-scale
X-ray, optical and radio structure implies a recent merger (see
section \ref{sec:contour}).  However, whilst the AGN is underheating, it
is not clear that this is the result of an undersized SMBH.  For SMBH
accretion close to the Eddington rate, cavity power reaches an
observed maximum of $\sim10^{-2}L_{\mathrm{Edd}}$
(\citealt{Russell13}).  For H1821+643, the maximum cavity power should
then be $\sim5\times10^{45}\ergps$, or roughly an order of magnitude
higher than is observed, and comparable to the cooling luminosity.
AGN feedback in H1821+643 may simply be undergoing a period of underheating.



\section{Conclusions}

Using a deep, short frame-time \textit{Chandra} observation and a
detailed PSF simulation, we extracted the properties of the hot
atmosphere within $20\kpc$ of the low-redshift quasar H1821+643.  The
soft-band surface brightness increases sharply by a factor of $\sim30$
within the central $10\kpc$.  This cannot be explained by an
undersubtraction of the quasar PSF or scattered quasar light.  The
hard-band surface brightness profile, which is dominated by the quasar
emission, shows no significant residuals on the arcsec scales of the
PSF.  The observed level of pileup is inconsistent with an
underestimation of the soft band PSF flux by the required factor of
$>3$ and the soft-band excess is significantly extended compared to
the PSF.  The gas properties in the soft-band peak are instead
consistent with thermal ISM emission.  The gas temperature drops
sharply from $2\keV$ at a radius of $20\kpc$ to $<0.4\keV$ at
$10\kpc$.  The gas density rises rapidly by more than an order of
magnitude over this radial range with a gradient that is consistent
with hydrostatic compression of the hot atmosphere.  The metallicity
is particularly low here and, based on the ionization parameter, the gas has
likely been photo-ionized by the quasar emission.  We show that the sharp increase
in gas density and decrease in temperature are consistent and conclude
that the soft-band peak is due to thermal ISM emission.  The radiative
cooling time of the gas is only $12\pm1\Myr$ at a radius of $6.5\kpc$
and falls below the free fall time.  The hot atmosphere at the centre
of H1821+643 has therefore formed a cooling flow.  Under the
assumptions of a steady and spherical gas flow, we resolve the sonic
radius.

The measured mass deposition rate of up to $3000\Msunpyr$ can easily
supply the high star formation rate of $120^{+120}_{-60}\Msunpyr$ in
H1821+643 and the required accretion rate onto the quasar of
$\sim40\Msunpyr$.  Multi-wavelength observations of the massive
molecular gas reservoir and extensive nebula of ionized gas support the
interpretation of a cooling flow.  Whilst it is likely that Compton
cooling by the quasar radiation enhances gas cooling on kpc scales,
the quasar luminosity is at least an order of magnitude too faint to
cool the entire $20\kpc$ region of the cooling flow.  Instead, the AGN
appears to be underheating the cluster core by a factor of $\sim5$.
Furthermore, the energy input by the cavities appears to suppress gas
cooling on tens of kpc scales but has limited impact within the
rapidly cooling region within $10\kpc$.  It is possible that the SMBH
is undersized, likely due to a recent merger, but it may alternatively
be undergoing a period of underheating.

\section*{Acknowledgments}

We thank the reviewer for helpful comments that improved the paper.  HRR acknowledges support from an STFC Ernest Rutherford Fellowship and
an Anne McLaren Fellowship from the University of Nottingham.  PEJN was supported under NASA contract NAS8-03060.  TEB is supported by the Doctoral Training Centre in AI at the University of Nottingham.  WNB acknowledges support from \textit{Chandra} X-ray Center grant AR1-22006X and the Penn State Eberly Endowment.  LC is supported by an STFC studentship.  This research has made use of data from
the \textit{Chandra} X-ray Observatory and software provided by the \textit{Chandra}
X-ray Center (CXC).  Support for this work was provided by the National Aeronautics and Space Administration through \textit{Chandra} Award Number G09-20119X issued by the \textit{Chandra} X-ray Center, which is operated by the Smithsonian Astrophysical Observatory for and on behalf of the National Aeronautics Space Administration under contract NAS8-03060.  This work used observations obtained with \textit{XMM–Newton}, an ESA science mission funded by ESA Member States and USA (NASA).  Many of
the plots in this paper were made using the Veusz software, written by
Jeremy Sanders.  This research made use of Astropy, a community-developed core Python package for Astronomy (\citealt{Astropy22}).

\section*{Data availability}

The \textit{Chandra} data described in this work are available in the \textit{Chandra} data archive (https://cxc.harvard.edu/cda/).  Processed data products detailed in this paper will be made available on reasonable request to the author.

\bibliographystyle{mnras} 
\bibliography{refs.bib}

\bsp	
\label{lastpage}
\end{document}